\documentclass[twocolumn]{aastex63}

\usepackage{comment}
 \usepackage{hyperref}
\usepackage{amsmath}
\usepackage{xfrac,datetime,hyperref}

\newcommand{\chas}[1]{\textcolor{purple}{#1}}

\newcommand{\kms}{km s$^{-1}$}
 \received{\today:\currenttime}
\shorttitle{WHAM Point Source\,46}
 \shortauthors{Kulkarni, Lin, Beichman}
 
\graphicspath{{./}{Figures/}}

\begin{document}

\title{The origin of WHAM Point Source~46}

\correspondingauthor{C. Beichman}
 \email{chas@ipac.caltech.edu}

%\author{Aahalaad Aaron Aardvaark}\affiliation{}
%\author{Bahamas Barbados Barracuda}\affiliation{}
 %\author{Zanzibar Zeren Zebra \&\ friends}\affiliation{}

\author[0000-0001-5390-8563]{S.\ R.\ Kulkarni}\affiliation{Owens Valley Radio Observatory, California Institute of Technology 249-17, Pasadena, CA 91125, USA}

\author[0009-0001-7405-1228]{Zeren Lin}
\affiliation{Dept of Astronomy California Institute of Technology, 1200 E. California Blvd., Pasadena,CA 91125}

\author[0000-0002-5627-5471]{Charles Beichman}
\affiliation{NASA Exoplanet Science Institute, IPAC, Pasadena, CA 91125}
\affiliation{Jet Propulsion Laboratory, California Institute of Technology, Pasadena, CA 91109, USA}

\author[0000-0001-7301-5666]{Alex S.\ Hill}\affiliation{Department of Computer Science, Math, Physics, \& Statistics, University of British Columbia, Okanagan Campus, Kelowna, BC V1V 1V7, Canada}

\affiliation{Dominion Radio Astrophysical Observatory, Herzberg Astronomy \& Astrophysics Research Centre, National Research Council Canada, Kaleden, BC V0H 1K0 Canada}

\author[0009-0005-1425-3601]{Xihan Deng}
\affiliation{Dept of Astronomy California Institute of Technology, 1200 E. California Blvd., Pasadena,CA 91125}

\author[0009-0002-2268-7352]{Tryston Raecke}
\affiliation{Dept of Astronomy California Institute of Technology, 1200 E. California Blvd., Pasadena,CA 91125}

\author[0000-0003-2821-1750]{Mateusz Matuszewski}
\affiliation{Dept of Astronomy California Institute of Technology, 1200 E. California Blvd., Pasadena,CA 91125}

\author[0000-0001-5982-0060]{Drew M. Miles}
\affiliation{Dept of Astronomy California Institute of Technology, 1200 E. California Blvd., Pasadena,CA 91125}

\author{Marty Anderson}\affiliation{1084 Cherriebell Road Mississauga, ON. L5E 2R3, Canada}

\author[0000-0002-8650-1644]{D. Christopher Martin}
\affiliation{Dept of Astronomy California Institute of Technology, 1200 E. California Blvd., Pasadena,CA 91125}

 \begin{abstract}
The Wisconsin H$\alpha$ Mapper (WHAM) surveyed the entire Galactic
sky in H$\alpha$ ($\vert v_{\rm LSR}\vert \lesssim 100\, {\rm
km\,s^{-1}}$) to approximately 0.1\,Rayleigh (R), albeit with a
1-degree beam. %The resulting WHAM Sky Survey, along with large area
%imaging in [\ion{S}{2}] and [\ion{N}{2}], laid the foundation for Warm Ionized
%Medium (WIM) science.
\cite{rcm+05} reported ``point sources" which stood
out against the Galactic background in space and velocity.  Half
of the sources are associated with plausible planetary nebulae and
OB stars. \citeauthor{rcm+05} suggested sub dwarfs for one quarter of
the sources.  Here, we investigate one such source, WPS\,46, for
which \citeauthor{rcm+05} suggested the sub-dwarf PG\,0931+691 to
provide the source of ionization.  With the Keck Cosmic Web Imager
we found numerous nebular emission lines within the vicinity of
WPS\,46, but we failed to find H$\alpha$ emission in the arc-minute
vicinity of PG\,0931+691.   The line ratios (BPT diagram and [\ion{S}{2}]/H$\alpha$) combined with the morphology are more consistent with AGN or LI(N)ER-like ionization than with pure warm ionized medium or \ion{H}{2} region-like photoionization.
Separately, we offer compelling reasons to 
argue that PG\,0931+691 cannot be the source of ionizing power
for WPS\,46.   We suggest that WPS\,46 is associated with
an intermediate velocity complex (IVC) and that H$\alpha$ and
nebula emission may arise as a result of a shock.  We conclude
by outlining a plan of action of using SDSS's  Local
Volume Mapper along with deep narrow band imagery obtained
by amateur astronomers to explore and study the ionized sky on sub-degree scales, in general, and specifically  studies of IVC and high-velocity
complexes.
\\\\
 \end{abstract}

\section{Introduction}
	\label{sec:Introduction}

Using the Wisconsin H$\alpha$ Mapper (WHAM) sky survey \citep{hrt+03},
\cite{rcm+05} identified 86 enhancements  at intermediate and high
Galactic latitude ($\vert b\vert > 10^\circ$), over a velocity range
$\vert v_{\rm LSR}\vert \lesssim {\rm 100\,km\,s^{-1}}$ and with an angular
size less than or comparable to the 1-degree WHAM beam; here LSR
refers to the Local Standard of Rest velocity frame.  These
enhancements were termed WHAM Point Sources or WPS. The flux densities
in H$\alpha$ ranged from $10^{-11}\,{\rm erg\,cm^{-2}\,s^{-1}}$ to
$10^{-9}\,{\rm erg\,cm^{-2}\,s^{-1}}$.  About half of the sources
were close (on the sky) to cataloged planetary nebulae and massive
O and early B stars.  Fourteen of these were near hot evolved
low-mass stars (sub dwarfs) that had no previously reported nebulosity.
For the remaining 29 sources, the authors did not report
any association.

Here, we focus on WPS\,46, which was
associated with PG\,0931+691 by \cite{rcm+05}.
The nominal position of WPS\,46 is ${\rm \alpha=09^h\ 25.8^m}, {\rm
\delta=+69^\circ\, 05^m}$ (J2000) and the Galactic coordinates are
${\rm l^{II}=143.9^\circ}, {\rm  b^{II}=38.6^\circ}$.  The  H$\alpha$
emission of WPS\,46 peaks at ${\rm-52\,km\,s^{-1}}$ and the full-width
at half-maximum is 31\,km\,s$^{-1}$.  The authors noted that WPS\,46
was present in two adjacent beams, that is WPS\,46 is extended.
Integrating the H$\alpha$ brightness with the solid angle results
in a flux density of H$\alpha$ of $(5\pm 0.08)\times 10^{-11}\,{\rm
erg\,cm^{-2}\,s^{-1}}$.  PG\,0931+691 was discovered in the
Palomar-Green survey  and classified as an sdO star \citep{gsl86}.
It has a G magnitude of 16.7, a Gaia parallax of $1.72\pm 0.06\,$mas,
(distance, $d = 580\pm 19$\,pc) and is located at $l^{\rm
II}=143.54^\circ$ and $b^{\rm II}=39.52^\circ$. Note that the star
is offset by $\sim0.94^\circ$ from the nominal centroid of WPS\,46.

The paper is organized as follows.  In \S\ref{sec:WHAM} we review
the WHAM findings. With the Keck Cosmic Web Imager (KCWI;
\citealt{Martin2010,mmm+18}) we observed
the region around PG\,0931+691 and some locations within WPS\,46.
 Data and line analysis are presented in  \S\ref{sec:KCWI}.  We
found strong nebular emission within the WPS\,46 region, but virtually
no emission
in the vicinity of PG\,0931+691.
\S\ref{sec:What_Powers_WPS_46} is devoted to identifying the
ionization process, photo-ionization   and/or ionization due to shocks.
In  \S\ref{sec:IVC} we present circumstantial evidence
linking WPS\,46 to an intermediate velocity cloud (IVC). We argue that
WPS\,46 is post-shocked gas.  In \S\ref{sec:future} we suggest
that the WPS catalog
could be a good pointer to shocked intermediate cloud sites. 
We note two new developments: deep narrow-band imagery provided by
amateur astronomers and the recently commissioned Local Volume
Mapper of the Sloan Digital Sky Survey-V (SDSS-V). 
We conclude that these developments will advance the study of the ionized
sky at arc-minute resolution.

% \begin{figure}[htbp]   % WHAM_map.m
%  \plotone{WHAM_spectra_waterfall.png}
%   \caption{\small The WHAM H$\alpha$ spectra in the vicinity of
%   PG\,0931+691.  The legend refers to angular offset, in degrees,
%   between the direction to PG\,0931+691 and WHAM beams.  We believe
%   that the sharp feature at $v_{\rm LSR}=+25\,{\rm km\,s^{-1}}$ is
%   imperfectly subtracted geo-coronal emission.}
%  \label{fig:WHAM_spectra}
% \end{figure}

\section{WHAM data}
	\label{sec:WHAM}
	
\begin{figure}[htbp]   % WHAM_map.m
 \plotone{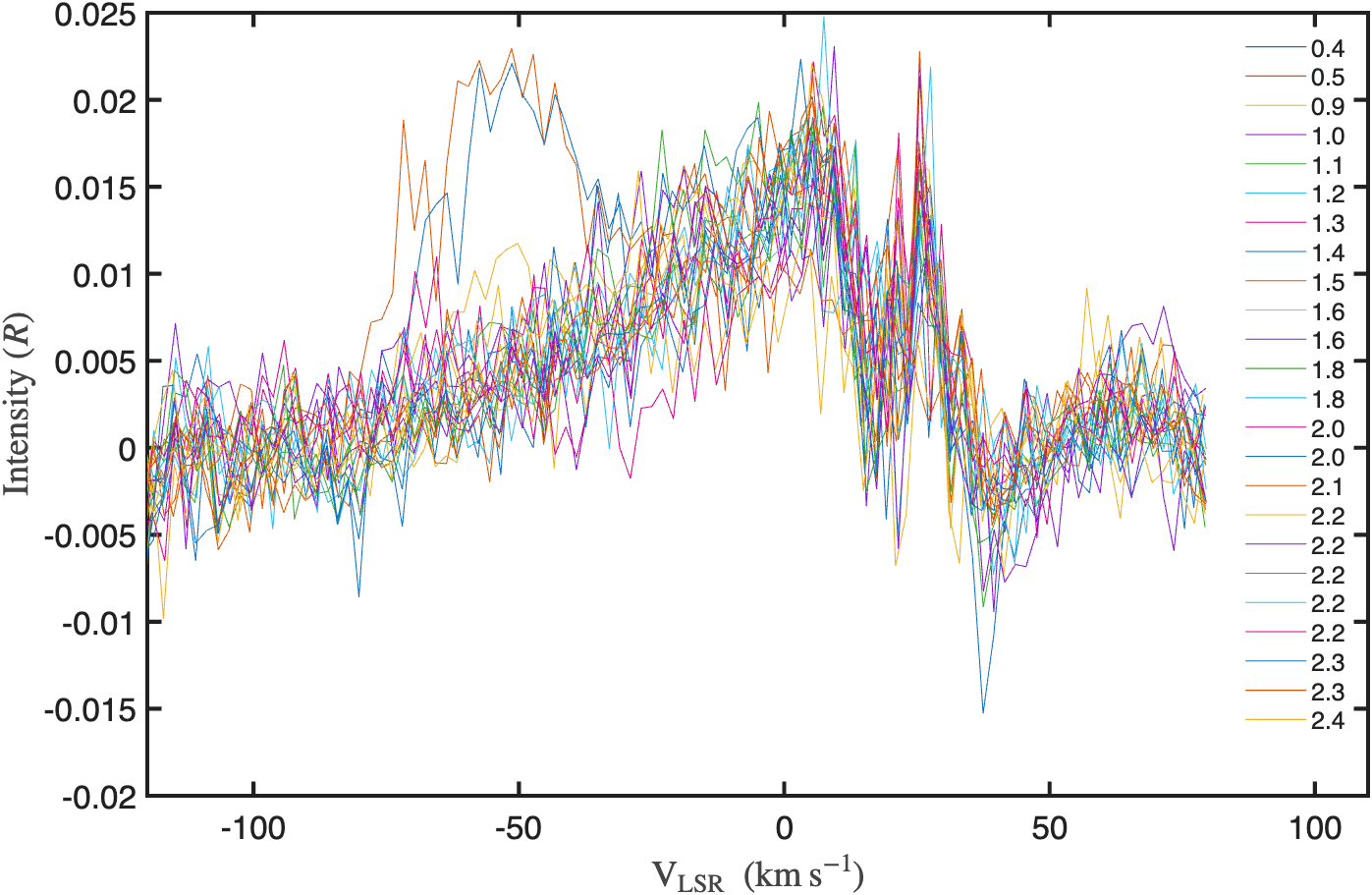}
  \caption{\small The WHAM H$\alpha$ spectra in the vicinity of
  PG\,0931+691.  The legend refers to angular offset, in degrees,
  between the direction to PG\,0931+691 and WHAM beams.  We believe
  that the sharp feature at $v_{\rm LSR}=+25\,{\rm km\,s^{-1}}$ is
  imperfectly subtracted geocoronal emission.}
 \label{fig:WHAM_spectra}
\end{figure}

The WHAM all-sky survey is described in \citet{hrt+03}. We
extracted\footnote{The data are hosted at
\url{https://www.astro.wisc.edu/research/research-areas/galactic-astronomy/wham/wham-sky-survey/wham-ss-data-release/}}
spectra within $2.5^\circ$ of PG\,0931+691. The 24 resulting spectra
are plotted in Figure~\ref{fig:WHAM_spectra}. In the velocity range,
$-75\,{\rm km\,s^{-1}} < v_{\rm LSR} < -35\,{\rm km\,s^{-1}}$, two
of the beams show a stronger emission relative to the other beams.
It is this excess emission that led \citet{rcm+05}
to the  discovery of WPS\,46.

%\section{WHAM data}
%	\label{sec:WHAM}

\begin{figure}[htbp]    % PG0931_WHAM.m
 \plotone{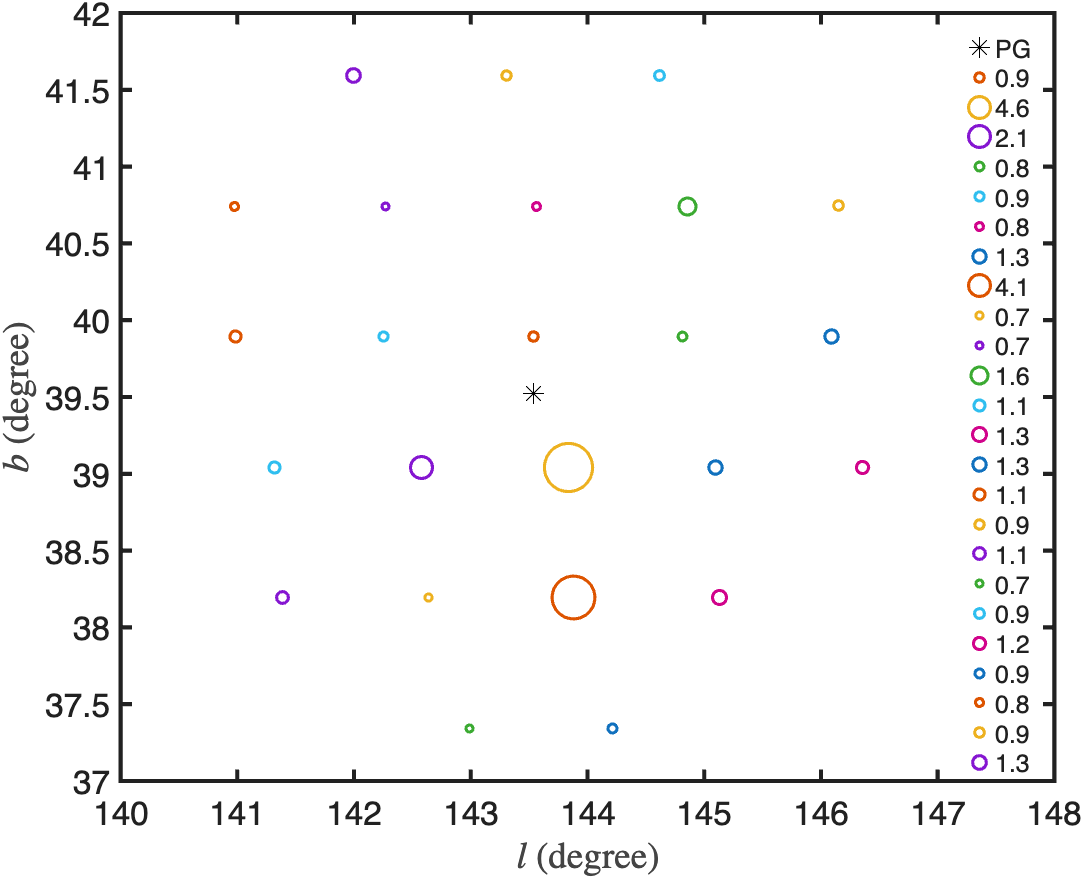}
  \caption{\small H$\alpha$ intensity, integrated over the velocity
  range $-75\,{\rm km\,s^{-1}} < v_{\rm LSR} < -35\,{\rm km\,s^{-1}}$,
  of the region around PG\,0931+691 (marked by ``$\textasteriskcentered$").
  Each beam is represented by a circle (marker) whose size represents the
  ratio of the emission in the direction normalized to the median
  intensity ($0.15\,R$) in this region of the sky.  The three
  brightest points have intensity of $0.69\,R, 0.61\,R$ and $0.31\,R$.
  }
 \label{fig:Halpha_map}
\end{figure}

A ``map" obtained by integrating the emission over the aforementioned
velocity range is displayed in Figure~\ref{fig:Halpha_map}.  In
this map PG\,0931+691 is marked by ``$\textasteriskcentered$".  From
this map it is clear that  {\it  (i)} WPS\,46 is offset
from PG\,0931+691, even at the coarse $1^\circ$ angular resolution
of WHAM and {\it (ii)} the source extends over two beams and perhaps
is present in a third beam also.

\section{KCWI observations}
	\label{sec:KCWI}

\begin{deluxetable}{lllll}
\tabletypesize{\scriptsize}
\tablecaption{KCWI setup (Runs 1--3)}
\label{tab:KCWI_setup}
\tablewidth{0pt}
\tablehead{
\colhead{\#} & 
\colhead{Slicer} &
\colhead{Gr(Red,Blue)} &
\colhead{$\lambda_c$\,(\AA)} &
\colhead{$R_S$} 
}
\startdata
1 & Large & RH1, BH3 & 6600, 4950 & 3250, 4500\\
2 & Medium & RH1, BH3 & 6600, 4950 & 6500, 9000\\
3 & Medium & RH4, BM & 9300, 3900 & 6500, 4000*
\enddata
 \tablecomments{Column 1 is the run number. 
 Run 1: UT March 25, 2025. 
 Run 2: UT December 18, 2025.
 Run 3: UT January 17, 2026.
 Column 2 specifies the choice of slicer.  Column 3 is the choice 
 of grating
 (first entry is for the red arm and second is for the blue arm).
 Column 4 is the central wavelength of the band (red arm
 followed by the blue arm).
 Column 5 is spectral resolution, $R_S=\lambda/\Delta\lambda$ for
 the red arm followed by blue arm.  Here,  $\Delta\lambda$ is the full-width
 half maximum of the line spread function. (*) The blue-arm spectral
 resolution was 4,000 for this run due to operational constraints
 that prevented us from modifying the instrument configuration;
 observations were therefore obtained with the medium-resolution
 grating.}
\end{deluxetable}

Observations were conducted with KCWI which is mounted on the right Nasmyth
port of the Keck~II telescope. KCWI is a two-armed (blue and red)
integral field unit (IFU) spectrograph that delivers spectra over
its field-of-view (FoV).  The FoV is set by the choice of the image
slicer, which also sets the width of the slit.  The smaller the slit width,
the higher the spectral resolution.  Three image slicers (``large'',
``medium", ``small") offer the following FoV:
$[33^{\prime\prime},16.5,8^{\prime\prime}]\times 20^{\prime\prime}$.
The corresponding pixels sizes are
$[1.35^{\prime\prime},0.69^{\prime\prime},0.34^{\prime\prime}]$.
The spectral resolution is determined by the width of the slit and
the choice of grating and can range from 1,000 to 20,000.

\begin{figure}
 \plotone{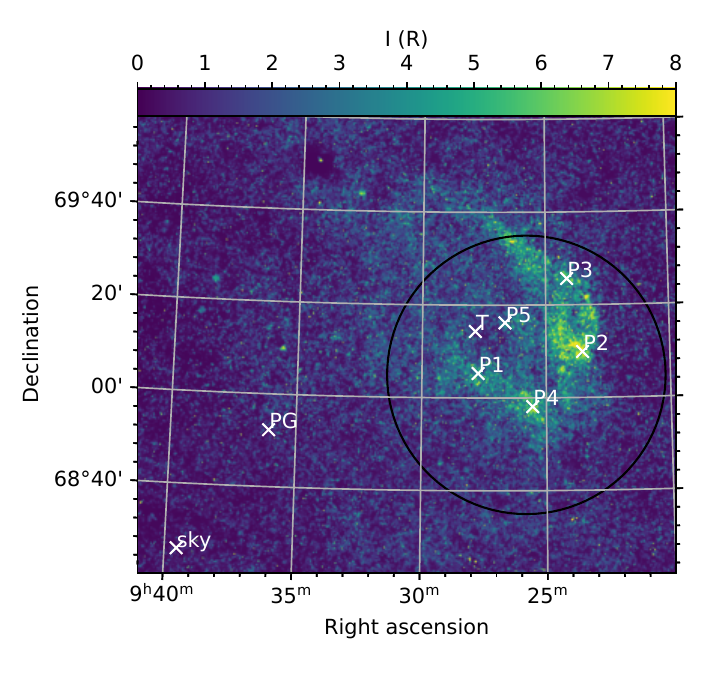} 
  \caption{H$\alpha$ image of the field of WPS\,46.  A circle with radius $0.5\arcdeg$
  is centered on the position of the WHAM pointing which contains
  WPS\,46.  ``PG" marks the pointing towards PG\,0931+691, ``T" is
  that for TYC~4376--968--1 and ``Sky" is the offset position to
  measure the sky devoid of H$\alpha$ light from WPS\,46.  The
  imaging data is from \citet{Z25}.} 
 \label{fig:KCWI_pointings}
\end{figure}

\begin{figure*}
 \plotone{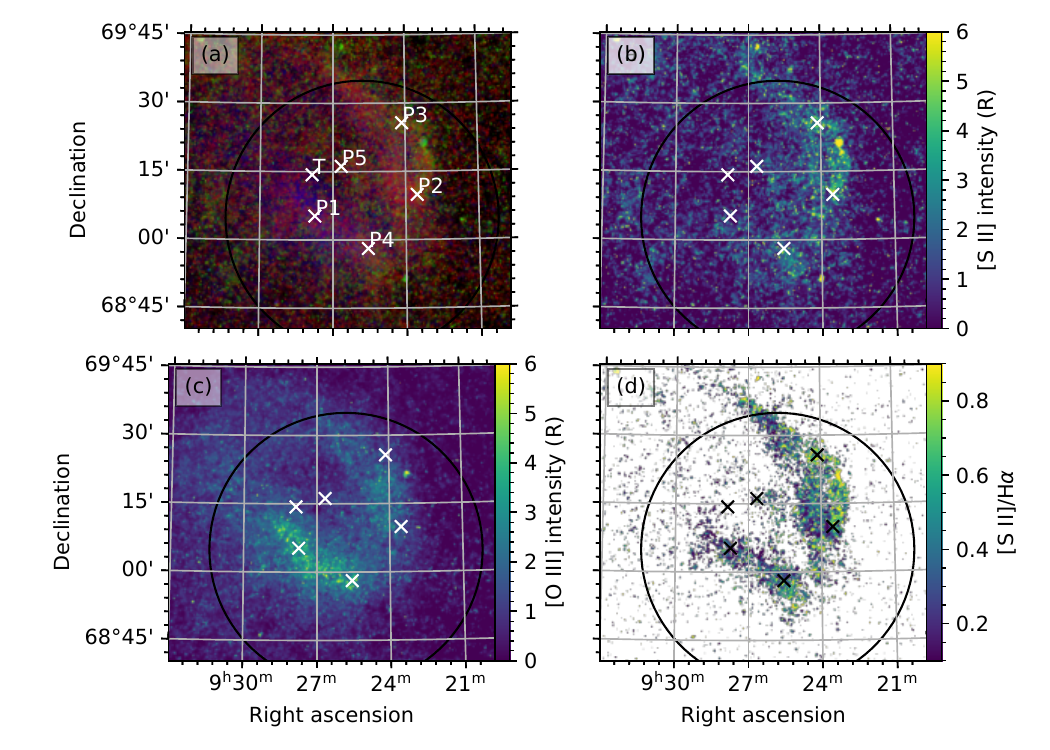}
  \caption{Composite (H$\alpha$ red, [\ion{S}{2}] ($\lambda6716$
  and $6731$) green, and [\ion{O}{3}]$\lambda 5007$ blue, panel $a$), [\ion{S}{2}] (panel $b$), [\ion{O}{3}] (panel $c$),
  and [\ion{S}{2}]/H$\alpha$ (in energy units, panel $d$) images of the region,
  using data from \citet{Z25} as in Figure~\ref{fig:KCWI_pointings}.
  The line ratio map is masked in directions for which
  $I_{\mathrm{H}\alpha} < 3 \textrm{ R}$. Labels in panel $a$ are as in Figure~\ref{fig:KCWI_pointings}. \label{fig:composite}}
\end{figure*}

We had three observing runs. The KCWI setup for the runs is summarized
in Table~\ref{tab:KCWI_setup}.  Contrary to normal usage, we use
KCWI in a ``light bucket" mode with the median sky spectrum (across
the FoV) serving as the primary observable.  Occasionally we refer
to each observation position as a ``beam".

\begin{deluxetable}{lrrr}
\tabletypesize{\scriptsize}
\tablecaption{Positions of beams: Runs 2 \&\ 3}
\label{tab:KCWI_runs_23_log}
\tablewidth{0pt}
\tablecolumns{4}
\tablehead{
 \colhead{Pointing} &
 \colhead{RA (J2000)} &
 \colhead{Dec (J2000)} &
 \colhead{E(B$-$V)}
}
\startdata
P1       & 09:27:44.62 & +69:05:16.8 & 0.116\\
P2        & 09:23:31.66 & +69:09:54.0 & 0.118\\
P3        & 09:24:08.14 & +69:25:40.0 & 0.213\\
P4        & 09:25:33.10 & +68:58:08.0 & 0.085\\
P5       & 09:26:39.34 & +69:16:11.0 & 0.213 \\
TYC~4376--968--1  & 09:27:51.34 & +69:14:20.0 & 0.167\\
PG\,0931+691        & 09:36:05.04 & +68:52:16.9 & 0.099\\
Offset~Sky        & 09:39:31.56 & +68:25:58.8 & 0.079 \\
\enddata
 \tablecomments{ Each entry corresponds to a single KCWI pointing
 (beam).  P1 through P5 refer to pointings at various bright points
 in WPS\,46 (see also Figure~\ref{fig:KCWI_pointings})  The Offset
 Sky pointing was used for sky subtraction and characterization of
 atmospheric emission. The integration time for each observation
 was 300\,s. E(B$-$V) in magnitude is from
 \href{https://irsa.ipac.caltech.edu/applications/DUST/}{\color{blue}IPAC/IRSA}.
 }
\end{deluxetable}

In the first run (UT March 25, 2025), we focused on positions in
the vicinity of PG\,0931+691.  Although the spectral resolution was
modest and the geocoronal H$\alpha$ emission was strong, emission at
the expected velocity offset of $-52\,\mathrm{km\,s^{-1}}$ would
have been spectrally resolved had it been present; no such emission
was detected.  The limits we placed were good enough to inform us
that the WHAM H$\alpha$ emission was not centered on PG\,0931+691
(see $\S$Appendix~\ref{sec:epoch1} for additional details of this
run).

After the first run, we became aware of the availability of narrow-band H$\alpha$ images
on ten arcsecond scales \citep{Z25}. This new H$\alpha$ image, shown in Figure~\ref{fig:KCWI_pointings}, showed that most of the ionized gas was in the West. 
The setup for the second  run (UT December 18, 2025) was informed
by lessons learned from the first run. We used the medium slicer.
The resulting higher spectral resolution improved the separation
of any velocity-offset emission from the geocoronal H$\alpha$ line. We shifted our observations to
several bright regions within WPS\,46, TYC~4376--968--1, 
and re-observed PG\,0931+91. We also observed a position to determine the
sky H$\alpha$ emission, free of WPS\,46.  The sky positions of the
eight beams are shown graphically in Figures~\ref{fig:KCWI_pointings} and \ref{fig:composite}
and are also noted in Table~\ref{tab:KCWI_runs_23_log}.  As will be
discussed later in the paper, there is considerable  diffuse emission
(``cirrus" first reported by \citealt{S76}) towards WPS\,46.  To this
end, in Table~\ref{tab:KCWI_runs_23_log}, we provide $E(B-V)$ for
each pointing location.  A third run was undertaken on UT January
17, 2026 towards the same set of positions, but the grating setup
was changed to include  [\ion{O}{2}] and [\ion{S}{3}].

\subsection{Data reduction}

Data were reduced using the standard KCWI data reduction pipeline
(DRP)\footnote{\url{https://kcwi-drp.readthedocs.io/en/latest/}},
which performs bias subtraction, flat-fielding, wavelength calibration,
and flux calibration. Additional cosmic-ray rejection for the
red-channel data was carried out using \texttt{DeepCR}
\citep{zhang2020deepcr}. Each reduced three-dimensional data cube
was collapsed into a one-dimensional ``light-bucket'' spectrum by
taking the median flux in each wavelength plane across the field
of view.  Uncertainties on the light-bucket spectra were estimated
via Monte Carlo resampling.  At each wavelength plane, 400 realizations
were generated by perturbing each spaxel with Gaussian noise drawn
from the DRP per-spaxel uncertainty cube.  The spatial median was
computed for each realization, and the 1-$\sigma$ uncertainty on
the median spectrum was taken as the standard deviation of the
resulting distribution of median values.

\begin{deluxetable}{lcc}
\tabletypesize{\scriptsize}
\tablecaption{Velocity reference-frame corrections}
\label{tab:velocity_corr}
\tablewidth{0pt}
\tablehead{
\colhead{Pointing} &
\colhead{$v_\mathrm{bary}$} &
\colhead{$v_{\mathrm{LSR,add}}$} \\ [-0.3cm]
\colhead{} &
\colhead{(km\,s$^{-1}$)} &
\colhead{(km\,s$^{-1}$)}
}
\startdata
\multicolumn{3}{c}{\textit{Run 2: 2025-12-18}} \\
P1          &  9.6 & 3.2 \\
P2          &  9.5 & 3.1 \\
P3          &  9.4 & 3.2 \\
P4          &  9.6 & 3.1 \\
P5          &  9.5 & 3.2 \\
TYC~4376--968--1    &  9.6 & 3.2 \\
PG\,0931+691         &  9.9 & 3.3 \\
Sky          & 10.2 & 3.3 \\
\hline
\multicolumn{3}{c}{\textit{Run 3: 2026-01-17}} \\
P1          & $-0.1$ & 3.2 \\
P2          & $-0.3$ & 3.1 \\
P3          & $-0.3$ & 3.2 \\
P4          & $-0.1$ & 3.1 \\
P5          & $-0.2$ & 3.2 \\
TYC~4376--968--1    & $-0.1$ & 3.2 \\
PG\,0931+691         &  0.3  & 3.3 \\
Sky          &  0.6  & 3.3 \\
\enddata
 \tablecomments{ $v_{\rm bary}$ is the barycentric velocity
 correction computed using \texttt{astropy}.  $v_{\mathrm{LSR,add}}$
 is the line-of-sight projection of the solar motion relative to
 the Local Standard of Rest; adding it to $v_\mathrm{bary}$ yields
 an approximate correction to the LSR frame.  } 
\end{deluxetable}

Following the initial DRP wavelength solution, the wavelength
calibration was refined using sky- and atmospheric features appropriate
to each spectral region.  For the red-channel observations (H$\alpha$,
[\ion{N}{2}], [\ion{S}{2}], and [\ion{S}{3}]), the wavelength calibration was anchored
to the OH night-sky emission lines \citep{osterbrock1996night,ofb97}.
In the blue channel, the wavelength solution for H$\beta$ and [\ion{O}{3}I]
was calibrated using the geocoronal H$\beta$ line.  There are no
strong lines in the vicinity of the [\ion{O}{2}]\,$\lambda\lambda$3727,3729.
Following \citet{hanuschik2003flux} we achieved wavelength calibration
by cross-correlating with a high-resolution night-sky atlas.  Velocity
corrections for barycentric and LSR motions were applied on a
per-pointing basis.  These are given in Table~\ref{tab:velocity_corr}.

\begin{deluxetable*}{lccccc}
\tabletypesize{\scriptsize}
\label{tab:kcwi_line_velocities}
\tablecaption{Emission-line centroid velocities}
\tablewidth{0pt}
\tablehead{
\colhead{Line} &
\colhead{P1} &
\colhead{P2} &
\colhead{P3} &
\colhead{P4} &
\colhead{P5}  \\ [-0.25cm]
\colhead{($\mathrm{\AA}$)} &
\colhead{(km s$^{-1}$)} &
\colhead{(km s$^{-1}$)} &
\colhead{(km s$^{-1}$)} &
\colhead{(km s$^{-1}$)} &
\colhead{(km s$^{-1}$)} 
}
\startdata
{[\ion{O}{2}] $\lambda3726$}   & $-47.7 \pm 4.5$ & $-47.0 \pm 0.9$ & $-41.3 \pm 1.2$ & $-49.5 \pm 1.4$ & $-46.7 \pm 5.7$ \\
{[\ion{O}{2}] $\lambda3729$}   & $-47.6 \pm 4.5$ & $-47.0 \pm 0.9$ & $-41.3 \pm 1.2$ & $-49.4 \pm 1.4$ & $-46.7 \pm 5.7$ \\
H$\beta$                         & $-52.1 \pm 1.6$ & $-49.3 \pm 0.8$ & $-46.1 \pm 1.5$ & $-52.7 \pm 1.2$ & $-46.8 \pm 3.8$ \\
{[\ion{O}{3}] $\lambda4959$}  & $-48.7 \pm 0.5$ & $-48.5 \pm 1.2$ & $-44.3 \pm 0.9$ & $-49.7 \pm 0.6$ & $-47.7 \pm 1.6$ \\
{[\ion{O}{3}] $\lambda5007$}  & $-47.0 \pm 0.2$ & $-46.3 \pm 0.4$ & $-43.4 \pm 0.3$ & $-48.5 \pm 0.2$ & $-46.5 \pm 0.6$ \\
H$\alpha$                        & $-53.9 \pm 0.2$ & $-51.5 \pm 0.1$ & $-48.4 \pm 0.2$ & $-55.3 \pm 0.2$ & $-53.1 \pm 0.4$ \\
{[\ion{N}{2}] $\lambda6548$}   & $-53.3 \pm 2.0$ & $-51.1 \pm 0.2$ & $-48.2 \pm 0.3$ & $-53.6 \pm 0.4$ & $-53.2 \pm 1.1$ \\
{[\ion{N}{2}] $\lambda6583$}   & $-55.2 \pm 0.5$ & $-52.2 \pm 0.1$ & $-48.7 \pm 0.1$ & $-54.9 \pm 0.1$ & $-54.6 \pm 0.3$ \\
{[\ion{S}{2}] $\lambda6731$}   & $-60.0 \pm 1.7$ & $-51.3 \pm 0.2$ & $-45.5 \pm 0.3$ & $-52.9 \pm 0.5$ & $-53.1 \pm 1.8$ \\ \hline
\chas{[\ion{S}{3}] $\lambda9069$}$^*$  & $-33.7 \pm 1.1$ & $-36.0 \pm 0.8$ & $-34.8 \pm 0.5$ & $-39.4 \pm 0.4$ & $-39.8 \pm 0.9$ \\
\chas{[\ion{S}{3}] $\lambda9531$}$^*$  & $-36.7 \pm 0.5$ & $-36.5 \pm 0.4$ & $-33.0 \pm 0.5$ & $-41.0 \pm 0.5$ & \nodata               
\enddata
 \tablecomments{Velocities are measured from Gaussian centroid fits
 to the sky-subtracted light-bucket spectra.  Quoted uncertainties
 are 1-$\sigma$ statistical errors from the centroid fits and reflect
 the flux uncertainties propagated through the fitting procedure.
 They do not include systematic uncertainties from the absolute
 wavelength calibration, which we estimate to be of order 1 to
 2\,km\,s$^{-1}$, based on comparisons with sky lines and calibration
 residuals.  For the [\ion{O}{2}]~$\lambda\lambda$3727,3729 doublet, the
 fitted full width at half maximum is typically ${\rm FWHM \simeq
 100\,km\,s^{-1}}$ (see Table~\ref{tab:KCWI_setup}), while for the
 remaining nebular lines the widths are ${\rm FWHM \simeq
 30\,km\,s^{-1}}$. These widths are close to the instrumental
 resolution (see Table~\ref{tab:KCWI_setup}).  $^*$The
  velocities of [\ion{S}{3}] lines are contaminated by adjacent telluric OH lines.}
\end{deluxetable*}

\begin{figure}
 \centering
  \includegraphics[width=0.5\textwidth]{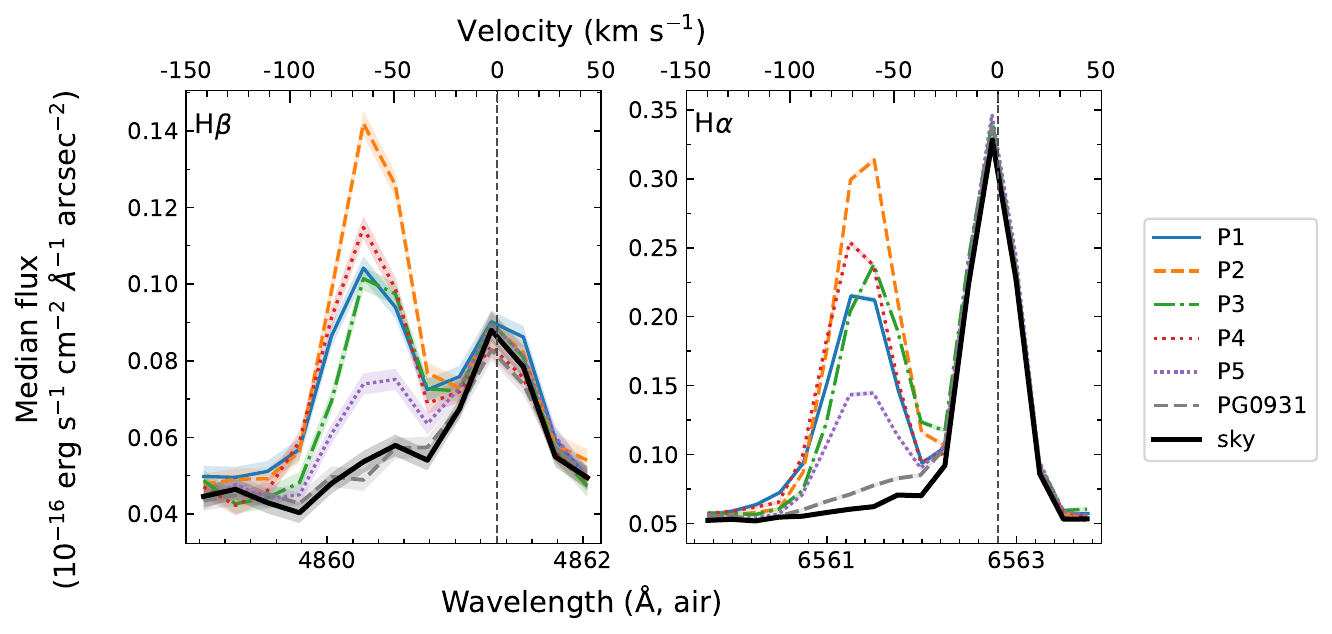}
   \caption{\small KCWI light-bucket spectra  (without sky subtraction)
   around H$\beta$ (left) and H$\alpha$ (right), shown in velocity
   space relative to the rest-frame air wavelengths.  Shaded regions
   indicate the 1-$\sigma$ uncertainties on the median spectra.
   Velocities are topocentric and so H$\alpha$ and H$\beta$ are
   centered at 0\,km\,s$^{-1}$. The geocoronal emission is clearly
   separated from the nebular emission; the geocoronal line widths
   are consistent with the instrumental resolution.}
 \label{fig:geo_Hb_Ha}
\end{figure}

As demonstrated in Figure~\ref{fig:geo_Hb_Ha}, the geocoronal
emission of H$\beta$ and H$\alpha$ is well separated in velocity
from the nebular emission.  The observed nebular emission lines are
unresolved or marginally resolved, with line widths consistent with
being resolution-limited in both the blue and red channels. The
centroid velocity measurements for all detected emission lines are
reported in Table~\ref{tab:kcwi_line_velocities}.

\begin{figure*}[htbp]
 \centering
  \includegraphics[width=0.9\textwidth]{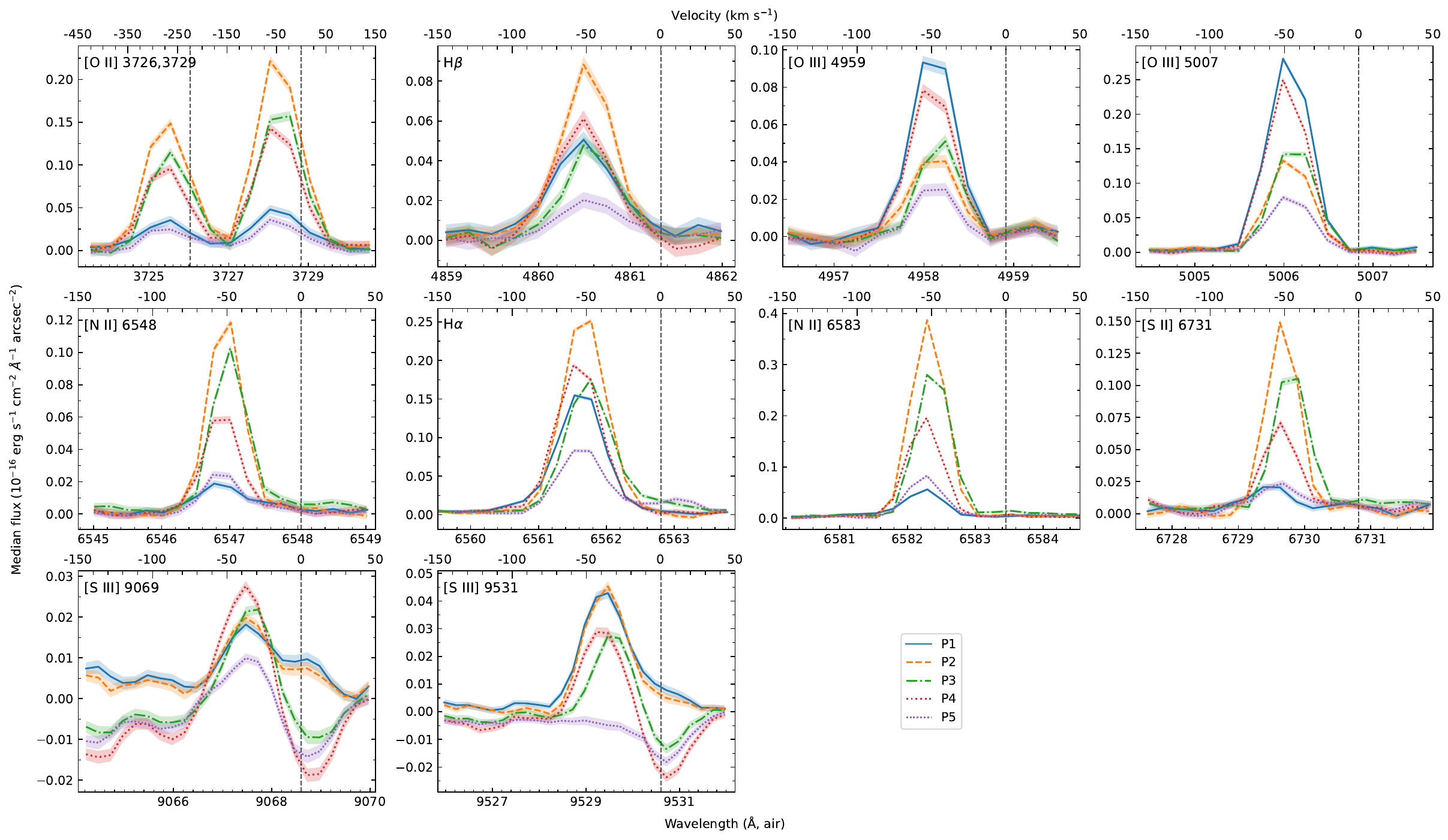}
  \caption{\small Sky-subtracted KCWI light-bucket spectra for all
  pointings, showing the full set of detected nebular emission lines
  across the blue and red channels from runs 2 and 3. Each
  panel displays a velocity-zoomed view of an individual emission
  line, with spectra from different pointings overplotted. Shaded
  regions indicate the 1-$\sigma$ uncertainties of the median
  spectra, derived via Monte Carlo resampling and shown in the
  corresponding color for each pointing. Vertical dashed lines mark
  the expected rest-frame wavelengths.}
 \label{fig:SummedSpectrum}
\end{figure*}

Representative light-bucket spectra showing H$\alpha$, H$\beta$,
and multiple nebular emission lines at each pointing are presented
in Figure~\ref{fig:SummedSpectrum}, with measured line fluxes
summarized in Table~\ref{tab:kcwi_line_fluxes_R}.

\begin{deluxetable*}{lcccccc}
\tabletypesize{\scriptsize}
\tablecaption{Integrated emission-line surface brightnesses \label{tab:kcwi_line_fluxes_R}}
\tablewidth{0pt}
\tablehead{
\colhead{Line} &
\colhead{P1} &
\colhead{P2} &
\colhead{P3} &
\colhead{P4} &
\colhead{P5} &
\colhead{${\rm A(\lambda)/A_V}$} \\[-0.25cm]
\colhead{($\AA$)} &
\colhead{(R)} &
\colhead{(R)} &
\colhead{(R)} &
\colhead{(R)} &
\colhead{(R)} &
\colhead{mag}
}
\startdata
{[\ion{O}{2}] $\lambda3726$} & $0.46 \pm 0.09$ & $2.05 \pm 0.09$ & $1.60 \pm 0.09$ & $1.33 \pm 0.09$ & $0.43 \pm 0.10$ & 1.54 \\
{[\ion{O}{2}] $\lambda3729$} & $0.67 \pm 0.10$ & $3.05 \pm 0.10$ & $2.32 \pm 0.09$ & $1.95 \pm 0.09$ & $0.54 \pm 0.10$ & 1.54 \\
H$\beta$                        & $0.48 \pm 0.05$ & $0.76 \pm 0.04$ & $0.44 \pm 0.04$ & $0.63 \pm 0.05$ & $0.20 \pm 0.05$ & 1.16\\
{[\ion{O}{3}] $\lambda4959$}& $0.81 \pm 0.03$ & $0.35 \pm 0.03$ & $0.39 \pm 0.03$ & $0.67 \pm 0.03$ & $0.22 \pm 0.03$  & 1.12\\
{[\ion{O}{3}] $\lambda5007$}& $2.26 \pm 0.03$ & $1.10 \pm 0.03$ & $1.24 \pm 0.03$ & $1.97 \pm 0.03$ & $0.68 \pm 0.03$ & 1.12\\
H$\alpha$                      & $2.28 \pm 0.04$ & $3.62 \pm 0.04$ & $2.42 \pm 0.04$ & $2.79 \pm 0.04$ & $1.13 \pm 0.03$ & 0.82 \\
{[\ion{N}{2}] $\lambda6548$} & $0.24 \pm 0.04$ & $1.37 \pm 0.03$ & $1.05 \pm 0.03$ & $0.71 \pm 0.03$ & $0.28 \pm 0.03$ & 0.82 \\
{[\ion{N}{2}] $\lambda6583$} & $0.61 \pm 0.03$ & $4.29 \pm 0.03$ & $3.19 \pm 0.03$ & $2.17 \pm 0.03$ & $0.88 \pm 0.03$ & 0.81 \\
{[\ion{S}{2}] $\lambda6731$} & $0.22 \pm 0.03$ & $1.63 \pm 0.03$ & $1.22 \pm 0.03$ & $0.76 \pm 0.03$ & $0.28 \pm 0.04$ & 0.79\\
{[\ion{S}{3}] $\lambda9069$}& $0.43 \pm 0.04$ & $0.44 \pm 0.03$ & $0.73 \pm 0.03$ & $1.05 \pm 0.03$ & $0.48 \pm 0.03$& 0.47 \\
{[\ion{S}{3}] $\lambda9531$}& $1.29 \pm 0.04$ & $1.28 \pm 0.04$ & $0.73 \pm 0.03$ & $1.01 \pm 0.04$ &  \nodata                 & 0.44
\enddata
 \tablecomments{ Values are integrated emission-line surface
 brightnesses in Rayleigh ($R$), reported with 1-$\sigma$ statistical
 uncertainties derived from the Gaussian flux measurements.  The
 [\ion{S}{3}]~$\lambda9531$ line is strongly affected by OH sky emission
 and is therefore less reliable than [\ion{S}{3}]~$\lambda9069$; for
 Pointing~5, where the feature is not securely detected, no integrated
 flux is reported.  The last column is ${\rm A(\lambda)/A_V}$ which
 was computed using
 \href{https://www.dougwelch.org/Acurve.html}{\color{blue}``Doug's
 Excellent Absorption Law Calculator"} (based on \citealt{ccm89}
 and assumes ${\rm R_V=3.1}$) hosted by IPAC.}
\end{deluxetable*}

\subsection{PG\,0931+691}
	\label{sec:PG0931}
	
As can be seen in Figure~\ref{fig:geo_Hb_Ha}, little or no emission
($<0.06\,R$) is detected at the position of PG\,0931+691 even at
the higher spectral resolution.

\subsection{Dust and Balmer Decrement}
\label{sec:dust}

We evaluate the impact of dust extinction using the Balmer decrement.
The Galactic foreground reddening toward WPS\,46, derived from the
SFD dust maps \citep{schlegel1998maps} with the recalibration of
\citet{schlafly2011measuring}, is $E(B-V)_{\rm MW} = 0.08$--0.21
across the five pointings.  Assuming the Milky Way extinction law
of \citet{ccm89} with $R_V = 3.1$ and an intrinsic Case~B ratio
${\rm (H\alpha/H\beta)_0 = 2.86}$, the predicted foreground Balmer
decrement is ${\rm (H\alpha/H\beta)_{MW} = 3.1}$--3.5.

The observed ratios, ${\rm (H\alpha/H\beta)_{\rm obs} = 3.3}$ --4.3,
are broadly consistent with the foreground prediction, with inferred
excess reddening at $\lesssim 1.8\sigma$ (Table~\ref{tab:balmer}).
We therefore find no compelling evidence for substantial internal
reddening beyond the Galactic foreground.

\begin{deluxetable*}{ccccccc}
\tabletypesize{\scriptsize}
\tablecaption{Balmer Decrement and Milky Way Dust Comparison \label{tab:balmer}}
\tablewidth{0pt}
\tablehead{
\colhead{Pointing} &
\colhead{$(H\alpha/H\beta)_{\rm obs}$} &
\colhead{$(H\alpha/H\beta)_{\rm MW}$} &
\colhead{$E(B-V)_{\rm Balmer}$} &
\colhead{$E(B-V)_{\rm MW}$} &
\colhead{$E(B-V)_{\rm int}$} &
\colhead{$E(B-V)_{\rm int}/\sigma$}
}
\startdata
P1 & $3.62 \pm 0.38$ & 3.20 & $0.24 \pm 0.11$ & 0.114 & +0.123 & 1.17 \\
P2 & $3.51 \pm 0.20$ & 3.21 & $0.21 \pm 0.06$ & 0.115 & +0.091 & 1.60 \\
P3 & $4.21 \pm 0.42$ & 3.53 & $0.39 \pm 0.10$ & 0.214 & +0.177 & 1.76 \\
P4 & $3.30 \pm 0.25$ & 3.10 & $0.15 \pm 0.08$ & 0.081 & +0.065 & 0.84 \\
P5 & $4.30 \pm 1.06$ & 3.53 & $0.41 \pm 0.25$ & 0.213 & +0.200 & 0.81 \\
\enddata
\tablecomments{
 Column (2) lists the observed Balmer decrement.  Column (3) gives
 the value predicted from Milky Way foreground reddening assuming
 the \citet{ccm89} extinction law with $R_V=3.1$ and Case~B
 recombination.  Column (4) is the color excess inferred from the
 observed Balmer ratio, while Column (5) is the foreground value
 derived from the SFD dust maps \citep{schlegel1998maps} with the
 recalibration of \citet{schlafly2011measuring}. Column (6) shows
 the inferred internal color excess, ${\rm E(B-V)_{\rm int} = E(B-V)_{\rm
 Balmer} - E(B-V)_{\rm MW}}$, and Column (7) gives its statistical
 significance.  All differences are $\lesssim 2\sigma$, indicating
 no compelling evidence for substantial internal reddening beyond
 the Galactic foreground.  } 
\end{deluxetable*}

\subsection{Line Velocity Distribution}
	\label{sec:velocity}

\begin{figure}[htbp]
 \centering
  \includegraphics[width=0.45\textwidth]{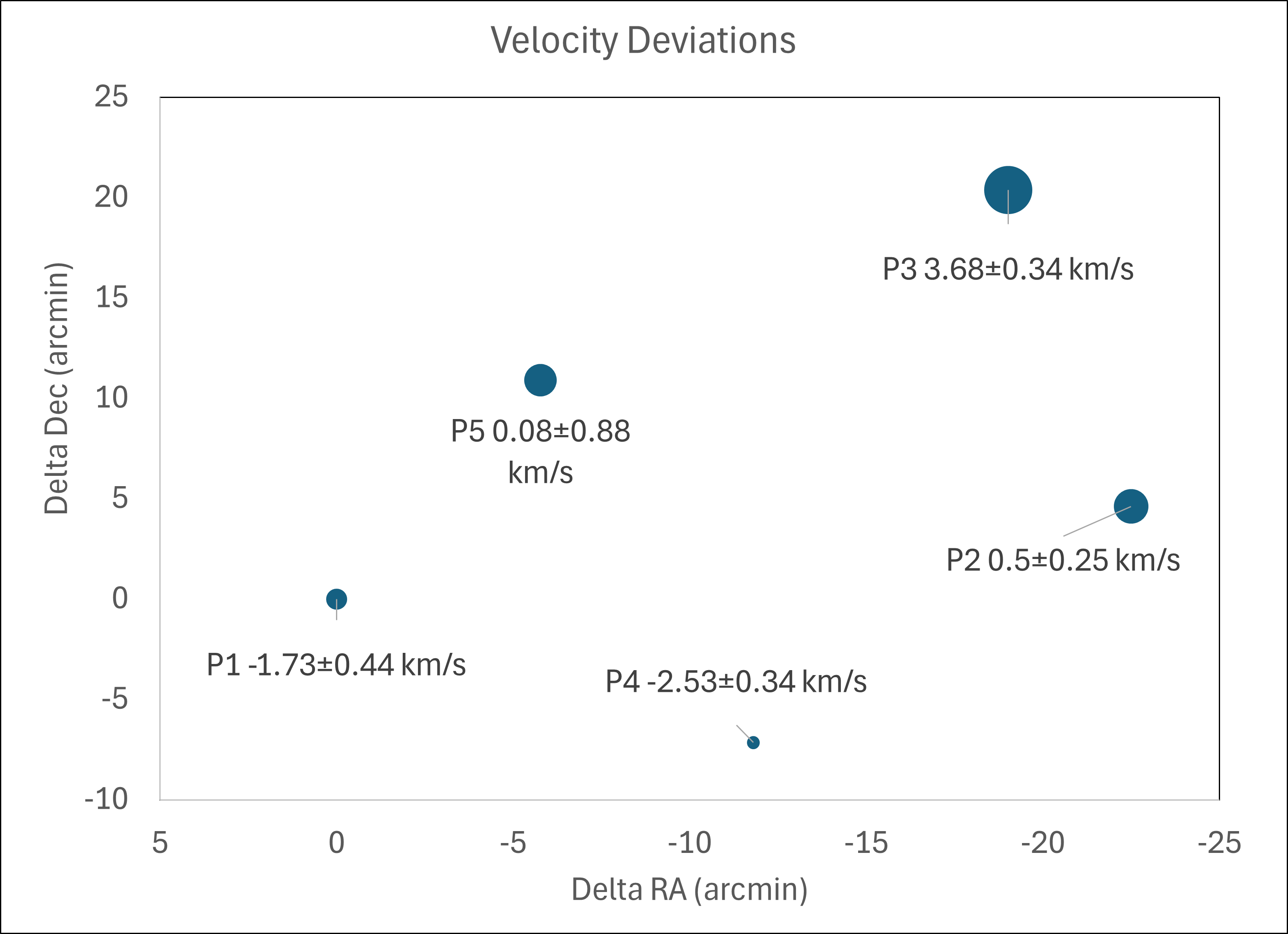}
   \caption{\small The average velocity and dispersion at each of
   the five  positions with respect to the position P1 averaged
   over   the four strongest lines with the most precise centroids
   (H$\alpha$, H$\beta$, [\ion{O}{3}]$\lambda$5007 and [\ion{N}{2}]$\lambda$6583;
   Table~\ref{tab:kcwi_line_velocities}).  The size of the symbol
   is proportional to the deviation in centroid velocity relative
   to the average over all positions, There are significant
   line-of-sight velocity differences within the nebula, up to 6
   \kms between P1 and P3. }
 \label{fig:VelDiff}
\end{figure}

The centroid velocities given in Table~\ref{tab:kcwi_line_velocities}
are broadly consistent with the WPS\,46 velocity of v$_{\rm LSR}=-52$
 \kms\ with significant emission extending from $[-6,60]\,$\kms
(\S\ref{sec:What_Powers_WPS_46}). The two long wavelength lines of
[\ion{S}{3}] 9069 \& 9531 at 36 \kms appear to be contaminated by nearby
atmospheric OH emission. The average centroid velocity for the other
wavelength lines across positions P1 to P5  is 49.6 \kms with a
dispersion of 2.7 \kms. Figure~\ref{fig:VelDiff} shows that the
centroid velocities vary significantly at different positions within
the nebula.

\begin{figure*}[htbp]
 \centering
  \includegraphics[width=0.7\textwidth]{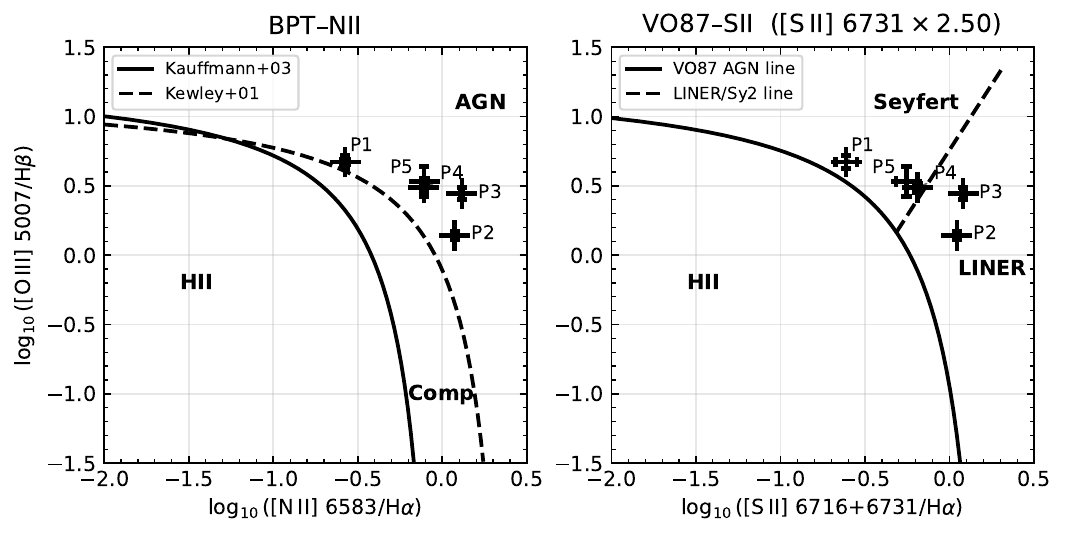}
    \caption{\small Emission-line diagnostic diagrams constructed
    from KCWI-blue and KCWI-red line ratios. Left: Standard BPT--\ion{N}{2}
    diagram showing $\log([\mathrm{O\,III}]\lambda5007/\mathrm{H}\beta)$
    versus $\log([\mathrm{N\,II}]\lambda6583/\mathrm{H}\alpha)$,
    with the empirical demarcation of \citet{kauffmann2003host}
    and the theoretical maximum starburst line of \citet{kewley2001theoretical}.
    Right: VO87--\ion{S}{2} diagram showing
    $\log([\mathrm{O\,III}]\lambda5007/\mathrm{H}\beta)$ versus
    $\log([\mathrm{S\,II}](\lambda6716+\lambda6731)/\mathrm{H}\alpha)$,
    using the AGN and LINER/Sy2 separation lines from
    \citet{veilleux1987spectral}.
    Because [\ion{S}{2}]~$\lambda6716$ is
    strongly affected by sky residuals, the total [\ion{S}{2}] flux is
    estimated assuming the low-density limit,
    $[\mathrm{S\,II}]\lambda6716/\lambda6731 = 1.496$. } 
 \label{fig:BPT_VO87}
\end{figure*}

\subsection{Line ratios}
	\label{sec:excitation}

In Figure~\ref{fig:BPT_VO87}  we display our measured line ratios
as standard BPT line ratio diagrams \citep{Baldwin1981}. KCWI
data place emission line ratios within WPS\,46 outside the typical
HII-region photoionization regime in the region associated with
active galactic nuclei (AGN) or low-ionization (nebular) emission region (LI(N)ER). These line ratios may be  due to  non-equilibrium ionization or to shock-excited emission  \citep{McCallum:2024,Zhou2025}. As described below, we prefer a shock-excitation model for most regions sampled within WPS~46.

\subsection{Spatially Resolved Line Ratios}
\label{sec:spatial_ratios}

The spatial distribution of the excitation provides additional
diagnostic leverage. Using the narrow band maps of
\citet{Z25}, we place the data on the KCWI surface-brightness
scale by comparing aperture-averaged fluxes at the five KCWI
pointings (Figure~\ref{fig:composite}). Circular apertures of radius $10\arcsec$, chosen to
match the KCWI field of view, are used to extract median
surface brightnesses in H$\alpha$, [O\,III]~$\lambda5007$, and
[\ion{S}{2}]~$\lambda6731$. For each line, a robust median-based
normalization factor, weighted by the KCWI S/N,
is applied to place the narrow band data on the KCWI flux scale.

\begin{figure}
 \plotone{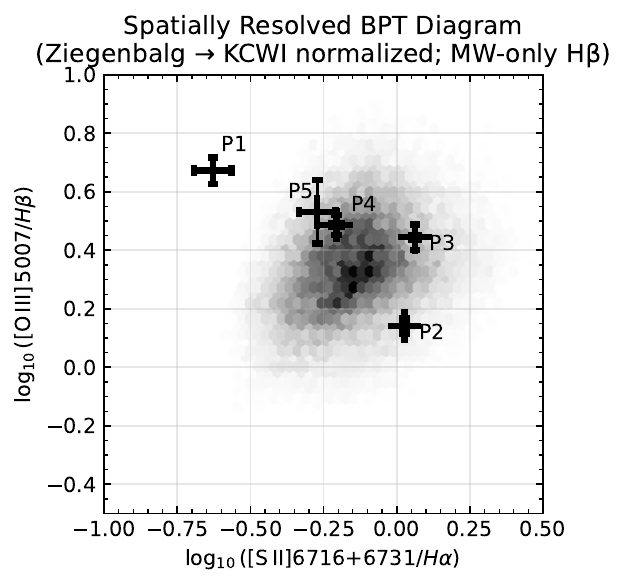}
  \caption{Spatially resolved BPT diagram for the KCWI field.  The
  grayscale hexbin shows the pixel density (arbitrary units) in
  $\log_{10}([\mathrm{S\,II}]\,6716{+}6731/\mathrm{H}\alpha)$ versus
  $\log_{10}([\mathrm{O\,III}]\,5007/\mathrm{H}\beta)$ space.  Pixels
  are included if they satisfy S/N $>3$ in H$\alpha$,
  [\ion{O}{3}]~$\lambda$5007, and [\ion{S}{2}]~$\lambda$6731 based on a robust
  background noise estimate, followed by morphological cleaning to
  remove isolated pixels.  The five KCWI pointings (P1--P5) are
  overplotted as black $+$ symbols with measurement uncertainties.}
 \label{fig:spatial_bpt}
\end{figure}

After normalization, we construct spatially resolved diagnostic
diagrams in the $\log([\mathrm{S\,II}]\lambda6731/\mathrm{H}\alpha)$–
$\log([\mathrm{O\,III}]\lambda5007/\mathrm{H}\beta)$ plane
(Figure~\ref{fig:spatial_bpt}). For consistency with standard
excitation diagnostics, H$\beta$ is inferred from H$\alpha$ assuming
Case~B recombination, $(H\alpha/H\beta)_0 = 2.86$, and correcting
only for the extinction of the Milky Way foreground as described above.  Only
pixels with S/N $>3$ in H$\alpha$, [O \, III]~$\lambda5007$, and
[\ion{S}{2}]~$\lambda6731$ are included, where the noise is estimated
using a robust background statistic; isolated detections are removed
by morphological filtering.

\subsection{Comparison to WIM diagnostics}
    \label{sec:WIM}
\newcommand{\nii}{\ensuremath{\textrm{[\ion{N}{2}]}\lambda6583}}
\newcommand{\siilow}{\ensuremath{\textrm{[\ion{S}{2}]}\lambda6716}}
\newcommand{\siihigh}{\ensuremath{\textrm{[\ion{S}{2}]}\lambda6731}}
\newcommand{\halpha}{\ensuremath{\textrm{H}\alpha}}

\begin{figure}
 \plotone{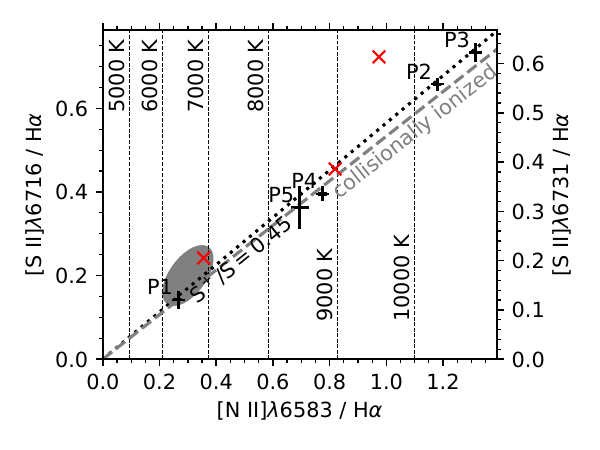}
  \caption{$\siilow/\halpha$ line ratio (derived from \siihigh\ as
  described in the text, with $\siihigh/\halpha$ indicated on the
  right) as a function of $\nii/\halpha$ for the five positions
  observed with KWCI. Temperatures and implied sulfur ionization
  states derived assuming photoionization following \citet{Madsen2006} (assuming N$^+/$N$ =
  0.8$ and H$^+/$H$=1.0$) are indicated. Red $\times$ signs indicate
  the line ratios for models C, D, and E (in order of increasing
  line ratios) of \citet{sm79}. The black dotted line shows the
  locus of S$^+/$S$=0.45$ calculated following \citet{Madsen2006},
  the gray ellipse approximates the region where \ion{H}{2} regions
  lie observationally \citep{Madsen2006}, and the gray dashed line
  shows the transition between photoionization-dominated (upper
  left) and collisional ionization-dominated (lower right) regions
  \citep{McCallum:2024}.}
    \label{fig:siiha_niiha}
\end{figure}

A plot of the $\siilow/\textrm{H}\alpha$ line ratio as a function
of $\nii/\textrm{H}\alpha$ diagnoses the temperature and ionization
state of the Warm Ionized Medium (WIM) \citep{Rand1998,Haffner1999,Madsen2006}. We used
\siihigh\ observations, multiplied by the collision strength ratio
$\Omega_{6716} / \Omega_{6731} = 1.496$ \citep{D11}, and considered
this equivalent to \siilow\ to allow direct comparison with WHAM
observations in the literature. The results are shown in
Figure~\ref{fig:siiha_niiha}.

All five points are close to a single
line in this space, implying similar line ratios $\siilow/\nii$.
If the gas is photoionized, this is consistent with a sulfur
ionization fraction $\textrm{S}^+/\textrm{S} \approx 0.45$. However, the $\nii/\halpha$ and $\siilow/\halpha$ line ratios each vary considerably. Points P2 and P3 are on the edge of the nebula with enhanced $\halpha$ intensity as well as an enhanced $\siilow/\halpha$ line ratio evident in the \citet{Z25} map (Fig.~\ref{fig:composite}). These two points also have the highest line ratios in Figure~\ref{fig:siiha_niiha}, with $\siilow/\halpha \approx 1.2$. Meanwhile points P4 and P5, which are in other parts of the cloud which are bright in $\halpha$ but not on the leading edge (Fig.~\ref{fig:composite}) have appreciably lower line ratios, $\siilow/\halpha \approx 0.7$. This line ratio is typical of the diffuse WIM \citep{Madsen2006,Haffner2009}. P1 is both the faintest in $\halpha$, in a diffuse area surrounded by bright emission (Fig.~\ref{fig:composite}), and has the lowest line ratio, $\siilow/\halpha \approx 0.3$. This line ratio is in the typical region occupied by \ion{H}{2} regions in WHAM observations \citep{Madsen2006,Haffner2009}.

All of these interpretations assume photoionization, but the line ratios are not consistent with the morphology. In particular, the faintest part of the region (P1) having an \ion{H}{2} region-like line ratio is not consistent with standard photoionization modelling. Because there is no local ionization source, photoionization could only arise due to the ambient ionizing radiation field. However that field would be fairly uniform, leading to consistent ionization ratios, which we do not observe. The line ratios are also consistent with shocks of $\sim 80 - 100
\textrm{ km s}^{-1}$ (models C, D, \& E of \citealt{sm79}, shown with red $\times$ signs in Fig.~\ref{fig:siiha_niiha}). A shock origin is more consistent with the morphology, especially with what appears to be the leading edge (points P2 and P3) having the highest implied temperature and highest shock velocity.

%\begin{figure}[htbp]
% \plotone{Figures/MadsenLineRatiowKCWI.png}
%  \caption{\small \color {red} Figure from \citep{Madsen2006} with KCWI points
%  superposed.  [\ion{S}{2}] intensity is multiplied by 2.5 to account for
%  missing [S II]\,$\lambda$6716. The [\ion{N}{2}]/H$\alpha$ and [\ion{S}{2}]/H$\alpha$
%  ratios vary considerably across the nebula, but are closely
%  correlated, suggesting regions of low ${\rm S^+/S}$ and low
%  temperature (25\% and 6,000 K) on the one hand and high  ${\rm
%  S^+/S}$ and high temperature ($>75$\% and 10,000 K) on the other.
% \label{fig:Madsen}}
%\end{figure}

\subsection{[\ion{O}{2}] doublet}

The strong [\ion{O}{2}]\,$\lambda\lambda$3726,3729 doublet with an energy
level of 3.3\,eV above ground is the primary coolant of hot
photo-ionized gas. Because of a large charge-exchange cross-section
and near coincidence of the ionization potential of atomic oxygen
and hydrogen, the ionization of oxygen tracks that of hydrogen. In
the WIM, [\ion{O}{2}]/H$\alpha$ proves to be an
excellent thermometer \citep{shr+00}; here, [\ion{O}{2}] is the sum of
the intensity of the doublet lines. For temperatures of $[6, 8, 10]\times
10^3\,$K, this ratio is  [0.11, 0.60, 1.72]. This can be compared
to the measured and extinction-corrected ratios for P1...P5: $[0.63\pm
0.12, 1.79 \pm 0.03, 2.51 \pm 0.04, 1.40 \pm 0.04,1.32 \pm 0.15]$.
In a photo-ionized framework the  temperatures exceed $10^4\,$K in
several regions of the nebula.

The [\ion{O}{2}] doublet intensity ratio is a traditional density indicator
\citep{of06}.  For electron density, $n_e\ll n_{\rm crit} \approx
10^3\,{\rm cm^{-3}}$, the ratio is 3/2, decreasing to 0.35 for
$n_e\gg n_{\rm crit}$.  Averaging the data for P1 through P4, we
find that this ratio is $1.47\pm 0.065$.  So all we can conclude is that
the plasma  is in the low density limit.

\section{What Powers WPS\,46}
	\label{sec:What_Powers_WPS_46}

In this section, we investigate the source of the ionization and
excitation of WPS\,46. We first summarize the key measurements of
WPS\,46 \citep{rcm+05}.  
The quoted flux density of H$\alpha$ is $f_{\rm H\alpha}=(5\pm
0.8)\times 10^{-11}\,{\rm erg\,cm^{-2}\,s^{-1}}$, which corresponds
to a photon flux density of $F_{\rm H\alpha}= 16.5\pm 2.6\,{\rm
phot\,cm^{-2}\,s^{-1}}$. The emission measure (EM) corresponding to the
peak H$\alpha$ intensity of $0.7\,R$ corresponds to an 
${\rm EM\approx 1.6\,cm^{-6}\,pc}$, assuming a nebular temperature of 8,000\,K.
The key physical parameter
that we are missing is $d$, the distance to WPS\,46. The luminosity
of H$\alpha$ photons is $2\times 10^{45}d_{\rm kpc}^2\,{\rm s^{-1}}$
where $d=d_{\rm kpc}\,{\rm kpc}$.

\begin{figure}[htbp] 
 \plotone{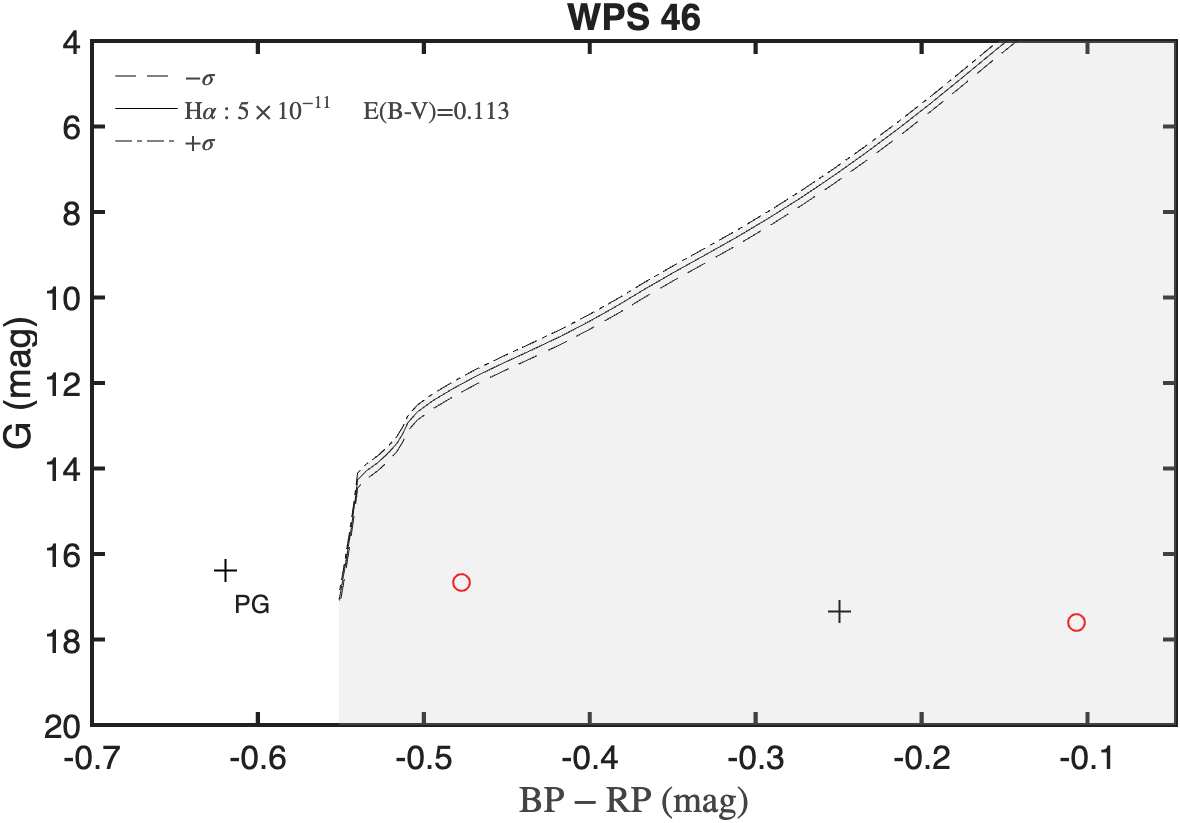}
  \caption{The Gaia CMD for hot stars within 1$^\circ$ of
  the formal position of WPS\,46. 
  The observed magnitudes is represented by red circles and the extinction corrected magnitudes
  by a ``+".
  Of the total of 12,734 stars
  only two stars, PG\,0931+691 (marked as ``PG") and GALEX J092257.6+694659 (a DA white dwarf with
  a parallax of 5.6\,ms)
  are hot and bright, ${\rm BP-RP<-0.2}$ and $G<20\,$mag. In a photo-ionization model  the ionizing source has to lie on or above the curves shown in the Figure. If the ionizing source is a black body then the color saturates at about $-0.55\,$mag. Points bluer than this could be due to incorrect extinction and deviation of the stellar SED from black body. Only PG\,0931+691 has the necessary ionizing flux to account for WPS\,46. }
 \label{fig:WPS_46_CMD} 
\end{figure}

The Gaia color–magnitude diagram (CMD) for hot stars within a degree of the formal position of
WPS\,46 is shown in  Figure~\ref{fig:WPS_46_CMD}.  
In \S\ref{sec:IonizingStarGaia} we consider a simple
model in which a star with black body spectrum (temperature, $T_*$)
is powering a Str\"omgren sphere. In equilibrium, the total rate of
recombinations (proportional to the H$\alpha$ luminosity) is balanced by
the ionizing luminosity, which is proportional to the Lyman continuum
stellar surface flux. The G band luminosity is also proportional to
the stellar flux, albeit at a lower frequency. It follows therefore
that the ratio of
the H$\alpha$ flux density to the G-band flux density should depend
only on $T_*$  (and without any dependence on radius and  distance
to the star). Thus, for a given flux density of H$\alpha$, there is a
curve for the flux density of the G-band that is solely a function of
$T_*$. The ionizing star must lie on this curve. In practice, 
the Gaia color, ${\rm BP-RP}$, can serve as a reasonable surrogate for
$T_*$. 

From Figure~\ref{fig:WPS_46_CMD} we see that only PG\,0931+691
has the necessary level of ionizing luminosity. [In  \S\ref{sec:PG}
we go beyond the simple black body model that forms the basis of
Figure~\ref{fig:WPS_46_CMD} and find that, to the level of precision
of interest here, the black body approximation suffices].

However, we exclude 
PG\,0931+691 as the source of ionization for a number of
reasons.  First, PG\,0931+691 is clearly offset from the bulk
of the nebula emission. 
The proper motion of PG\,0931+691 is $[\mu_\alpha, \mu_\delta] = [-1.2,-8.1]\,\textrm{ mas yr}^{-1}$,
which corresponds to $[-3,-22]\,{\rm km\,s^{-1}}$ along the right
ascension and declination axes, respectively. This low velocity does not favor any
model that calls upon the star's velocity to explain the angular offset.
Finally, the expected LSR velocity of the
gas in the vicinity of PG\,0931+691, given its parallax and Oort's
A-constant of ${\rm 15\,km\,s^{-1}\,kpc^{-1}}$, is only ${\rm
-8\,km\,s^{-1}}$, which is very different from the $-52\,{\rm
km\,s^{-1}}$ mean velocity of WPS\,46.

The next possibility is that WPS\,46 is a ``zombie" nebula, that
is, a nebula in which the ionizing star has faded away. This would
require that the cooling timescale of the star be smaller than the
recombination timescale. Of the species presented in
Table~\ref{tab:kcwi_line_fluxes_R} the recombination timescale for
O$^{++}$, $8.7\times 10^3n_e^{-1}\,$yr, is the shortest while that
for O$^+$ and H$^+$ is ten times longer. Given the inferred EM we
see that $n_e\gtrsim 0.1\,{\rm cm^{-3}}$ is reasonable. The corresponding recombination timescale is therefore of order
$10^5$ yr. Thus, we
do not favor the zombie hypothesis.

A third possibility is that WPS\,46  is illuminated by diffuse Lyman
continuum radiation, with observed H$\alpha$ emission arising from 
fluorescence.  Perhaps the best example of the fluorescence model 
is the emission of H$\alpha$ from high-speed clouds
(HVCs)\footnote{The edited compilation by  \cite{vws+04} is an
excellent starting point for readers interested in the phenomenology
of HVCs}; see \citealt{bm99,pbv+03,T04}.  \cite{bhw+12} present an
impressive set of deep observations of HVC~A, one of the most
extensive such complexes. The complex is located about 10\,kpc in
the halo of the Galaxy. The authors argue that the faint H$\alpha$
emission, $\lesssim 0.1\,R$, is powered by diffuse Lyman continuum
leaking from the Galactic disk. However, this mechanism cannot
explain the observed emission $\gtrsim 1\,R$ that we see in WPS\,46.

Having run out of photo-ionization models, we now explore the
possibility of ionization and excitation due to shocks, specifically
low-velocity shocks. Indeed, in the previous section, we found that
the BPT diagram analysis favors shock excited emission
(\S\ref{sec:excitation}) and the [\ion{S}{2}]/H$\alpha$ analysis 
(\S\ref{sec:WIM}) shows better agreement with low-velocity 
shocks than with the photo-ionization mechanism invoked for the WIM.

%Thus, direct evidence (line ratios) and circumstantial evidence (velocity) favor
%the low-velocity shock model.  In the
%next section, we look for a possible origin of low velocity shocks.

\section{An Intermediate Velocity Cloud Origin}
	\label{sec:IVC}

We start by noting that WPS\,46 is located in the second
quadrant and also at high latitude ($b_{\rm II}=38.6^\circ$) and
the mean LSR velocity of the H$\alpha$ emission is $-52\,{\rm km\,s^{-1}}$.
WPS\,46 is clearly not a part of the cold Galactic gas. This reasoning
led us to review the HI data at the 
Argelander Institut f\"ur Astronomie (AIfA) H~I Surveys Data
Server\footnote{\url{https://www.astro.uni-bonn.de/hisurvey/index.php}}. We 
retrieved an Effelsberg-Bonn H~I spectrum \citep{wkf+16} towards the
centroid of the WHAM nebula. To our pleasant surprise, we found
strong H~I emission  at about $-53\,{\rm km\,s^{-1}}$.

% Vizier for Bekhti et al. paper
% https://cdsarc.cds.unistra.fr/viz-bin/cat/J/A+A/594/A116
% The only way to get the data is from CDS (above).  
% The data comes in 20x20 deg^2 cubes

\begin{figure*}[htbp!]   %EffelsbergHImap.m:
 \centering
  \includegraphics[width=0.67\textwidth]{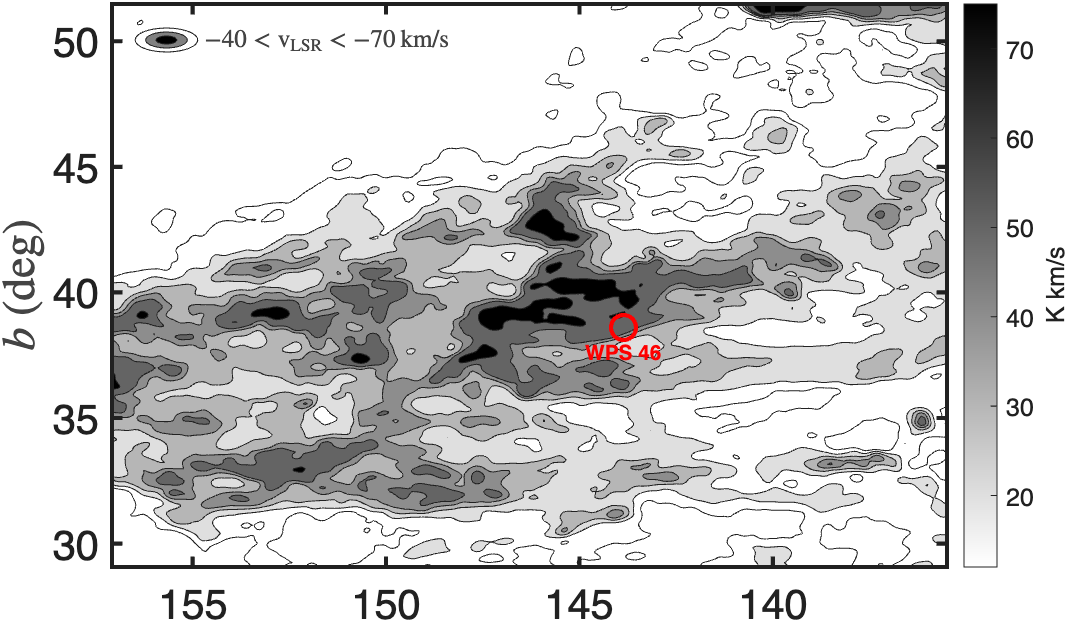} \vskip 15pt
   \caption{\small Effelsberg $\int T_s(v)dv$ in the
   vicinity of WPS\,46 with the integration limited to $-70 <
   v_{\rm LSR} < -40\,{\rm km\,s^{-1}}$; here $T_s$ is the spin temperature
   of H~I and $v$ is the radial velocity.  The position of WPS\,46
   is marked by a large red circle.}
 \label{fig:Effelsberg_WPS46}
\end{figure*}

As can be seen in Figure~\ref{fig:Effelsberg_WPS46}, WPS\,46
appears to be  on the outskirts of an IVC. This IVC is part of the
Intermediate Velocity Arches Complex (IV Arches) whose velocity
ranges from ${\rm -40\,km\,s^{-1}}$ to ${\rm -100\,km\,s^{-1}}$
\citep{kd96}. The distance to IV Arches is estimated to be between
1\,kpc to 2\,kpc \citep{kd96,sfk11}.  With little doubt, the gas in
the IV Arches, like other IVCs, is shocked. In support of this statement,
we note that it is peppered with molecular concentrations  \citep{rkl+16}
-- which are reasonably explained as a cooled and compressed post-shock
gas.

IVC ``K" and IVC ``L" have been mapped with WHAM \citep{hrt01,H05}.
In both cases, we see a rich image in H$\alpha$.  In contrast, the
H$\alpha$ image of WPS\,46 shown in Figure~\ref{fig:KCWI_pointings}
seems to be very simple. However, the reader should be aware that
the image is not deep.  

Fortunately, it came to our attention that
the region containing WPS\,46 apparently has been of some considerable
interest to ``amateur" astronomers. In Figure~\ref{fig:Vulcan} we
display a deep H$\alpha$ \&\ [\ion{O}{3}] image (obtained as described in Appendix $\S$\ref{sec:Vulcan}).  It  may
well be that the structure we see  in H$\alpha$ (long filaments,
short filaments, and various blobs) is part of the IVC shocked gas
complex.  Scattering of starlight by interstellar dust particles
gives rise to gray-cirrus features. In fact, this cirrus may well
be a parcel of shocked gas that has cooled and condensed.
We end this section by concluding that circumstantial evidence
links WPS\,46 as arising from post-shock gas in IV Arches.

%Simulations by \cite{swp+22} of the
%Pegasus-Pisces Arch Intermediate Velocity Complex
%(${\rm -50\,km\,s^{-1}}$, line-of-sight; 100\,km\,s$^{-1}$ true
%speed).

% The  scale is $5.93^{\prime\prime}$/pixel.

\begin{figure*}[htbp]
 \centering
  \includegraphics[width=0.8\textwidth]{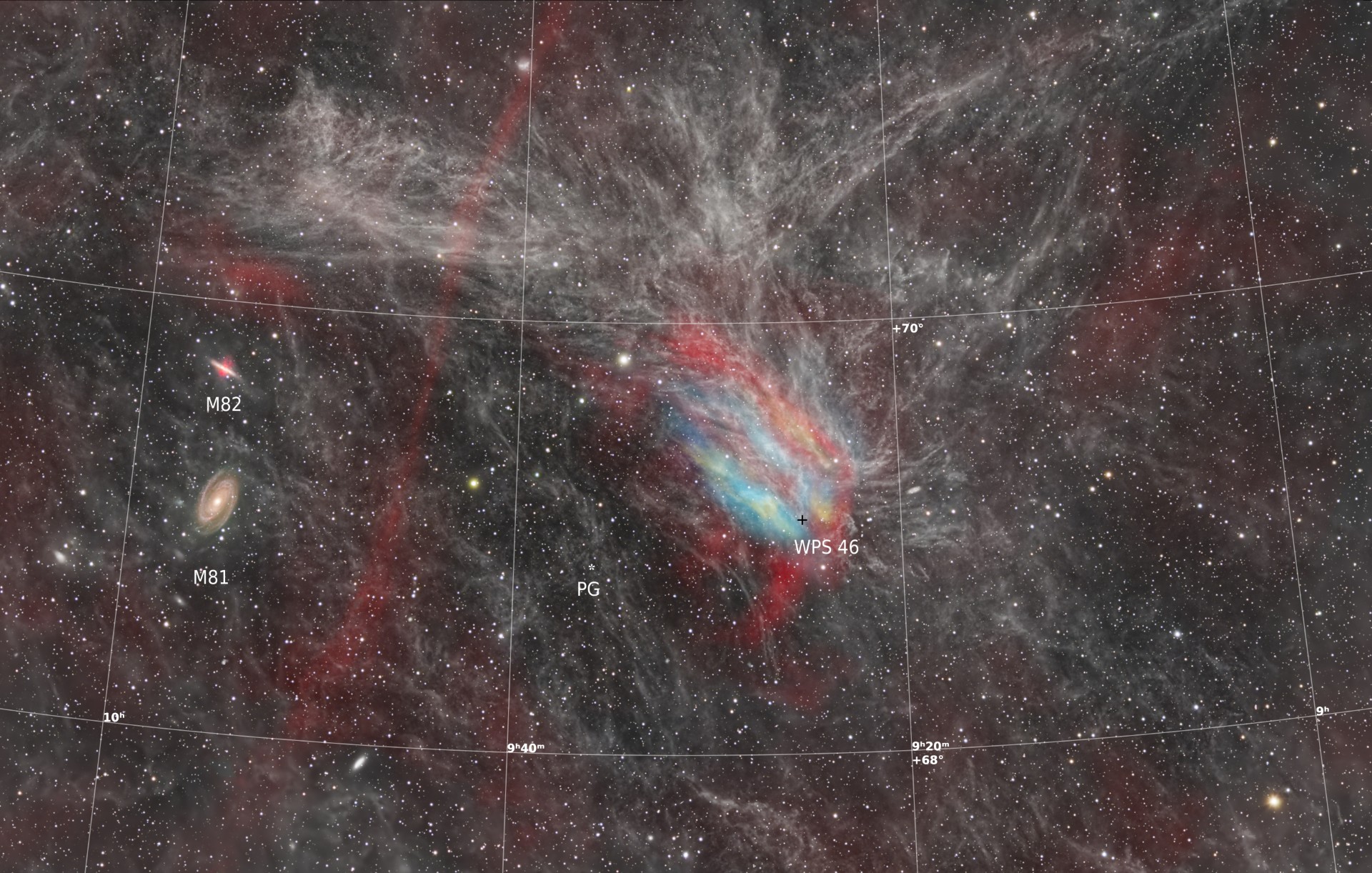}
   \caption{\small The ``Vulcan" nebula. North is up and East to
   the left.  The size of the image is about $6^\circ$ (RA) by
   $4^\circ$ (declination).
   The nominal position of WPS\,46 is marked by a ``+" sign while
   PG\,0631+691 is marked by ``*" (and annotated as PG). Color
   coding: red (H$\alpha$) and blue [\ion{O}{3}]. Gray is imaging in
   continuum bands and is an excellent tracer of dust. The grayish
   cirrus features are due to scattering by interstellar dust (see
   \citealt{S76}).  We draw the reader's attention to several diffuse
   H$\alpha$ blobs and also the curious slanted wavy thin H$\alpha$
   feature to the West of M81.  The technical details of the data
   and data reduction are summarized in \S\ref{sec:Vulcan}.}
 \label{fig:Vulcan}
\end{figure*}

\section{Prospects for Future Study \label{sec:future}}

The above discussion shows that the H$\alpha$ sky is likely to be
rich at the sub-Rayleigh brightness level with structures on
sub-degree angular scales.  As it so happens two new developments
-- deep narrow band imagery by amateur astronomers and the Local
Volume Mapper (LVM; \citealt{dbk+24}) of the Sloan Digital Sky Survey V (SDSS-V) --
make it timely to explore the diffuse ionized medium on sub-degree
angular scales. In the following, we summarize these two developments.\\

\subsection{ Deep narrow-band imagery from amateurs: A treasure trove.} We
start this discussion by acknowledging the key role of deep narrow
band images provided by amateur astronomers.  The \cite{Z25} survey 
has a fine angular resolution (say,
$10^{\prime\prime}$) compared to the 1$^\circ$ angular resolution
of WHAM and sufficient sensitivity to follow sources at the level of
a Rayleigh. The data from this survey were key in our discovery of the
nature of WPS\,46.

Next, the ultra-deep narrow-band image(s) of WPS\,46
(Figure~\ref{fig:Vulcan}) should be amazing even for professional
astronomers.  Not only is it aesthetically pleasing, but it provides
a dramatic picture of the collision of an HVC cloud with the
interstellar disk of the Galaxy.  Amateur astronomers refer to this
nebula as the Vulcan nebula, given its fiery appearance.  Professional
astronomers may wish to refer to the ``Great Galactic Splash".

We conclude that the resolution of the nature of the members of the
WPS catalog will proceed rapidly given the availability of the
\cite{Z25} catalog. Collaboration between professionals and
amateurs will likely result in dramatic pictures of astronomical
phenomenology (cf.\ Figure~\ref{fig:Vulcan}).\\

\subsection{The Local Volume Mapper: A powerful diagnostic of
HVCs and IVCs.} For diffuse spectral imaging, the millennium saw a
switch from Fabry-P\'erot spectrometers to massive IFU spectrometers.  The LVM
of the Sloan Digital Sky Survey (SDSS) Phase V has almost the same
``spectral grasp" as KCWI and MUSE, but its fractional square degree
FoV makes LVM very well suited for investigations of
sources in the WPS catalog as well as a reinvigorated study of IVCs
and HVCs.  In detail, the LVM has a  total of 1,801 fibers coupled
to a 16.1-cm aperture telescope. These fibers subtend a cone of
angular diameter $35.3^{\prime\prime}$ (``spaxel"). The on-sky solid
angle is 490 square arc-minutes. The fiber-fill fraction in the
focal plane is 83\%. Thus, the apparent solid angle on the sky is
about 590 square arc-minutes.

The fibers feed three spectrographs, each of which covers the wavelength
range 3600-9800\,\AA~\ at a spectral resolution of about 4,000.
This wide range allows observations of an impressive list of forbidden
lines: [\ion{O}{2}], [\ion{Ne}{3}], [\ion{O}{3}], [\ion{N}{1}], [\ion{O}{1}], [\ion{N}{2}], [\ion{S}{2}], and [\ion{S}{3}],
and also recombination lines of hydrogen and helium.  The stated
goal is the 5-$\sigma$ detection of $1\,R$ (Rayleigh), in each fiber,
in a channel centered on H$\alpha$, for 15 minutes of integration
time.  Averaging, say, $n=100$ spaxels should lead to a surface
brightness limit of $n^{-1/2}\,R=0.1\,R$ (with a solid angle of 27
\, square arc-minute, equivalent to $5^\prime\times 5^\prime$). At
this level of sensitivity it should be possible to start searching
for H$\alpha$ from the emission from IVCs and the brightest HVCs.

For now, assume observations at high latitude.  The spectral
resolution of the LVM is 4,000. This means that HVC emission ($\vert
v_{\rm LSR}\vert > 100\,{\rm km\,s^{-1}}$)  will be clearly resolved
from the geocoronal emission.  In contrast, IVC emission will show up
as blue or red tails to the geocoronal emission.  The barycentric
velocity of the geocoronal emission varies with the orbital phase
of the Earth by $\pm 30\cos(\beta)\,{\rm km\,s^{-1}}$ where $\beta$
is the ecliptic latitude. There are times of the year when the IVC
contributions can be cleanly separated from the geocoronal emission.
An observation strategy that optimizes this separation would be
very helpful for the study of IVCs. Finally,  dark time is needed
for sensitive observations in [\ion{O}{2}] and H$\alpha$, while observations
in [\ion{S}{3}] can be undertaken in gray time without a significant loss
in sensitivity.

We end this section by noting the power of Gaia CMD in identifying or ruling out putative sources of photo-ionization 
(cf.\ S\ref{sec:IonizingStarGaia}).

\section{Conclusions}
	\label{sec:Conclusions}

The WHAM sky survey is a cornerstone in our understanding of the WIM. WHAM
admittedly had a coarse angular resolution (1$^\circ$ beam), but
this was compensated by an unmatched sensitivity of $0.1\,R$ at a
spectral resolution of 25,000.

\citet{rcm+05} presented a list of high latitude ($\vert b\vert>10^\circ$)
WHAM ``point sources", sources with an angular size less than or
comparable to a single 1-degree beam of WHAM. The H$\alpha$ flux
limit is $10^{-11}\,{\rm erg\,cm^{-2}\,s^{-1}}$.  For some of these
sources, the authors suggested possible associations with hot white
dwarfs or sub dwarfs. One such source was WPS\,46 with a claimed
association with PG\,0931+691.  This star was initially classified
as a hot sub dwarf, but is now classified as a hot DO white dwarf.

Using KCWI in a light bucket mode,
we searched for and failed to find H$\alpha$ emission in the vicinity
of PG\,0931+691.  However, informed by recently available higher
angular narrow-band H$\alpha$ images \citep{Z25}, we changed our
pointing in subsequent KCWI runs and  detected strong
rich nebular line emission. The totality of the line data favors a
low-velocity shock origin for the gas in WPS\,46.

The Gaia parallax and the proper motion, PG\,0931+691, do not account
for the angular offset and the anomalous velocity of WPS\,46.  We
searched and failed to find other plausible ionizing stars. Instead,
we found that the WPS\,46 lies on the outskirts of the IVC, sharing the same $v_{\rm LSR}$ velocity.  This
complex is located between 1\,kpc and 2\,kpc away.  This finding 
provides a physical origin for the shock that we
inferred from the line data. 

According to \cite{lhm+22} most intermediate velocity complexes are
located at $\vert z\vert\lesssim 1.5\,$kpc and the coverage fraction
is very high,  $f_c\approx 0.9$. This high fraction means that WPS
sources with anomalous velocities are likely to arise in intermediate
velocity complexes.  Incidentally, the covering fraction for high velocity clouds
is $f_c\approx 0.14\ (\vert z\vert \lesssim 3\,$kpc) and $f_c\approx
0.6\ (\vert z\vert\lesssim 14\,$kpc). However, the expected emission
of H$\alpha$ from HVCs is $\lesssim 0.1\,R$, which is below the WPS
sensitivity limit.

The detailed study of WPS~46 presented here is a indicator of what a new generation of instrumentation can bring to the study of the Warm Interstellar Medium. The availability of the deep, narrow-band imaging obtained with  cameras on small telescopes combined with with spectroscopy from IFU spectrometers like KCWI and, eventually, with  SDSS's Local Volume Mapper (LVM) will extend  the study of the diffuse WIM far beyond the pioneering WHAM survey.

\facility{Keck:II (KCWI), Keck Observatory Archive (KOA)} 
\software{
\texttt{KCWI DRP} 
https://kcwi-drp.readthedocs.io/en/latest/ 
}

\begin{acknowledgements}

The team would like  to thank James D.\ Neill for his assistance with the KCWI pipeline. SRK thanks Dr.\ N.\ Reindl (Landessternwarte K\"onigstuhl, Heidelberg,
Germany), Dr.\ Peter Kalberla  (Radioastronomisches Institut der
Universit\"at Bonn, Germany) and Dr.\ H. Bond (Penn State University)
and Dr.\ Michael Shull (University of Colorado, Boulder) for consultations.
CAB thanks the Dominion Radio Astronomy Observatory for their
hospitality during the time when some of this work was being carried
out. ASH thanks Dr.\ R.\ A.\ Benjamin, Dr.\ L.\ McCallum, and Dr.\
K.\ W.\ Wood for useful discussions.

Part of the data analysis was performed at the Jet Propulsion
Laboratory, California Institute of Technology, under a contract
with the National Aeronautics and Space Administration (80NM0018D0004).

This research has made use of the Keck Observatory Archive (KOA), which is operated by the W. M. Keck Observatory and the NASA Exoplanet Science Institute (NExScI), under contract with the National Aeronautics and Space Administration.

The authors wish to recognize and acknowledge the very significant
cultural role and reverence that the summit of Maunakea has always
had within the Native Hawaiian community. We are most fortunate to
have the opportunity to conduct observations from this mountain.
This research has used the Keck Observatory Archive (KOA),
which is operated by the W. M. Keck Observatory and the NASA Exoplanet
Science Institute (NExScI), under contract with the National
Aeronautics and Space Administration.

\end{acknowledgements}

\bibliography{bibPG}{}
\bibliographystyle{aasjournal}

\appendix

\section{Initial Epoch of KCWI data}
\label{sec:epoch1}

\begin{figure}[htbp]   %KCWI_skyspec.m
 \centering
  \includegraphics[width=4in]{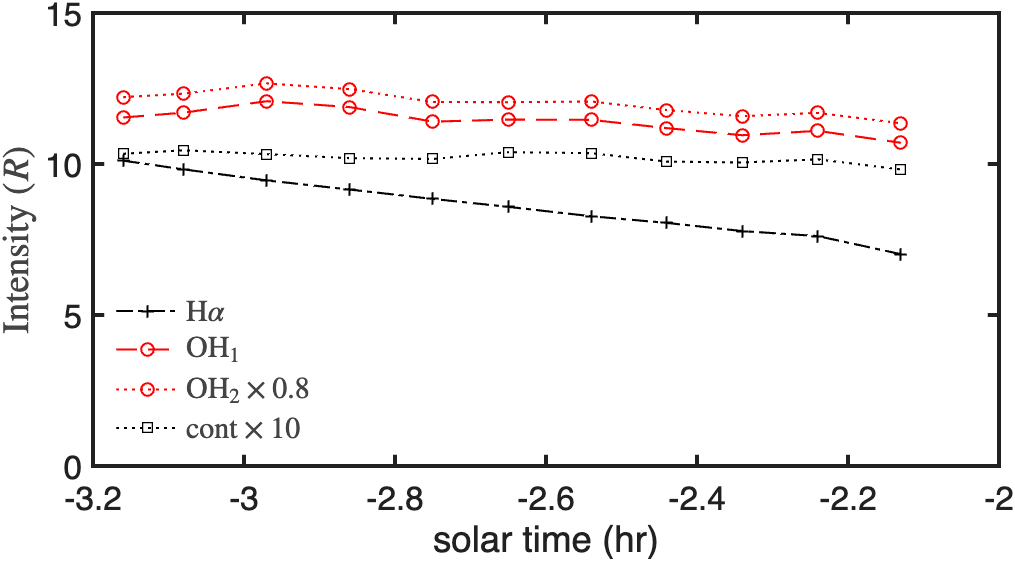}
   \caption{\small The run of the geocoronal 
   H$\alpha$ intensity,  the line integrated fluxes of OH lines (marked in
  Figure~\ref{fig:SummedSpectrum}) and the continuum  as a function
  of true solar time (see caption to Table~\ref{tab:KCWI_log2} for
  explanation). For the lines the 
  integration is over the entire line whereas for the continuum the integration is over 
  {1\,\AA}.
  Solar time of 0 hour corresponds to the sun at
  nadir.  To accommodate these four quantities on the same graph
 some of the quantities are multiplied by a factor (and noted in
  the legend).    }
 \label{fig:Beam_SkyEmission}
\end{figure}

The 2025-03-29 KCWI observations targeted PG\,0931+691 and its
vicinity.  No significant nebular emission was detected.
The spectra are dominated by geocoronal H$\alpha$
emission, which varies in intensity over a 3-hour duration of 
the observations (see Figure~\ref{fig:Beam_SkyEmission}).  

\begin{deluxetable}{cccccc}[htbt!]
\tablecaption{KCWI Observing Log for 2025-03-29\label{tab:KCWI_log2}}
\tablewidth{0pt}
\tablehead{
\colhead{\#} &
\colhead{$\delta\alpha$ ($^\prime$)} &
\colhead{$\delta\delta$ ($^\prime$)} &
\colhead{$\Delta T$ (hr)} &
\colhead{$\tau$ (s)} &
\colhead{$\sec(z)$}
}
\startdata
1  &  0.02  &  0.00  &  0.42  & 240 & 1.55 \\
1  &  0.02  &  0.00  &  0.42  & 240 & 1.55 \\
2  &  0.02  &  0.00  &  0.50  & 240 & 1.55 \\
3  &  0.48  &  0.00  &  0.61  & 300 & 1.54 \\
4  & $-0.45$&  0.00  &  0.72  & 300 & 1.54 \\
5  &  0.02  &  0.25  &  0.83  & 300 & 1.53 \\
6  &  0.02  & $-0.25$&  0.93  & 300 & 1.53 \\
7  &  0.48  &  0.25  &  1.04  & 300 & 1.52 \\
8  &  0.48  & $-0.25$&  1.14  & 300 & 1.52 \\
9  & $-0.45$&  0.25  &  1.24  & 300 & 1.52 \\
10 & $-0.45$& $-0.25$&  1.34  & 300 & 1.52 \\
\hline
11 & 19.08  & $-26.05$ & 1.45 & 300 & 1.51 \\
\enddata
 \tablecomments{ The first column is an index corresponding to the
 beam number referenced in the text.  Columns (2) and (3) give the
 offsets in right ascension and declination, respectively, in
 arcminutes relative to the position of PG\,0931+691.  Column (4)
 lists the time elapsed since UT 2025 March 29 06:00, column (5)
 gives the exposure time per pointing, and column (6) lists the
 airmass.  Beam~\#11 corresponds to the off-source sky pointing.
 On 2025 March 29, the average time of sunset occurred at 00:35
 HST.  The true solar time is measured relative to true solar
 midnight and is given by $\mathrm{UT} - 9.58~\mathrm{hr}$, where
 UT is expressed in hours.  } 
\end{deluxetable}

%The fluctuations around the median (per pixel) are about $\sigma=0.003$.%

As explained in the main body, after we became aware of the higher
resolution H$\alpha$ imagery by \cite{Z25} we renewed KCWI observations
but directed towards the main body of emission of WPS\,46.  Here
we focus on the properties of the night sky and geocoronal emission,
since, in our experience, the run proved to be instructive. As shown
in Figure~\ref{fig:Beam_SkyEmission} the continuum level adjacent
to the H$\alpha$ line  due to  is remarkably stable. The OH lines
arise in the mesosphere at an altitude of about 90\,km. In this
layer, the exothermic  reaction ${\rm O_3+H\rightarrow O_2+OH}$
leaves the OH radical in an excited state (+3.32\,eV). The excited
OH radical radiates across the optical and NIR bands (``Meinel"
bands). The temperature of the molecules is between 150\,K and
250\,K, so the line widths are small, 0.3\,km\,s$^{-1}$.  

In contrast,
H$\alpha$ is the result of fluorescence, primarily due to solar
Ly$\beta$ excitation of hydrogen atoms in the thermosphere (and
exosphere), well above the low altitude where OH lines arise..  As
the sun sets, the cylinder of shadow cast by the earth lengthens
and the geo-coronal H$\alpha$ emission gradually reduces and reaches
a minimum of about $2\,R$ to $4\,R$, depending on solar activity,
at midnight \citep{nmr+08}. Thus, the steady decline of H$\alpha$
emission is well understood. It is by no means clear to us why
there should be any physical correlation between geocoronal H$\alpha$
emission and the Meinel bands (cf.\ \citealt{zww+21}).

We make the following parenthetical remark.
The geocoronal H$\alpha$ is directly dependent 
on the ``activity" of the sun. Our first run as taken a year after the sun reached peak solar activity in October 2024. Geocoronal H$\alpha$ emission is expected to reach a minimal value in 2030.

\section{Constraining the ionizing star with Gaia data}
 \label{sec:IonizingStarGaia}
 
Consider a star with radius $R_*$ and blackbody temperature, $T_*$
immersed in a homogeneous medium of atomic hydrogen, number density,
$n_{\rm H}$.  In due course, a Str\"omgren sphere of radius $R$ is
formed with an electron (proton) density of $n_e(n_p)$.  In the
Str\"omgren ``radiation bound" approximation, all ionizing photons
are absorbed by the local nebula:
 \begin{eqnarray}
   \frac{4\pi}{3}R^3 \alpha n_en_p &=& 4\pi R_*^2 \pi q(T_*) \ \
   {\rm where\ } q=\int_{\nu_1}^\infty \frac{B_\nu(T_*)}{h\nu}d\nu
	\label{eq:recomb}
 \end{eqnarray}
where $\alpha$ is the total recombination coefficient rate (case A
or case B, as appropriate), $B_\nu(T_*)$ is the Planck blackbody
intensity function at frequency, $\nu$, and $\nu_1={\rm IP}/h$ with
IP being the ionization potential of hydrogen.  Let $F_{\rm H_\alpha}$
be the density of the H$\alpha$ photon flux as observed by an
observer at distance $d$. We then have
 \begin{equation}
	\frac{4\pi}{3}R^3 \alpha_{\rm H\alpha} n_e^2= 4\pi d^2
	F_{\rm H\alpha} \label{eq:Halpha}
 \end{equation}
where $\alpha_{\rm H\alpha}$ is the coefficient of recombination coefficient
for the production of H$\alpha$. We have set $n_e=n_p$.

Dividing Equation~\ref{eq:Halpha} by \ref{eq:recomb} by we get
$F_{\rm H\alpha}=\beta\pi (R_*/d)^2q(T_*)$ where $\beta=\alpha_{\rm
H\alpha}/\alpha$. The Gaia G-band flux density of the ionizing star
is given by $f_{\rm G}=\pi (R_*/d)^2 B_{\rm G}(T_*)$ where $B_{\rm
G}(T_*)$ is the Planck brightness in the G-band.  Dividing these
two equations, we find
 \begin{eqnarray}
   \frac{f_{\rm G}}{F_{\rm H\alpha}}&=& \frac{1}{\beta}\Bigg[\frac{B_{\rm
   G}(T_*)}{q(T_*)}\Bigg] = \frac{\mathcal{R}}{\beta} \ {\rm where}\
   \
	\mathcal{R} = \frac{x_{\rm G}/[\exp(x_{\rm G}-1]}{\int_{x_1}^\infty
	x^2/[\exp(x)-1]dx} \ .\label{eq:f(G)}
    \label{eq:f_G}
 \end{eqnarray}
Here, $x=h\nu/k_BT$, $x_1=h\nu_1/k_BT$, $x_{\rm G}=h\nu_{\rm
G}/k_BT$ with $\nu_G$ being the central frequency of the G band. $\beta$
slowly declines with $T$, the nebular temperature: 0.49 at T=5,000\,K
and decreases to 0.42 at T=20,000\,K (see \S14.2.3 of \citealt{D11}).
We set $\beta=0.45$.  Next, we assume that the wavelength centroid
of the Gaia G band is 622\,nm and the flux density 
corresponding to G=0 is
3229\,Jy.

% Filter, effective wavelength (nm), Vega_zero_mag [ erg / (Angstrom s cm2)]
% BP, 489.2, 4.05e-9  --> 3233 Jy and nu=6.127E14 Hz
% RP, 750.0, 1.29e-9
% G, 622 nm, 3229 Jy
% provided by Kareem El-Badry

Once $\beta$ and $\nu_G$ are set, the ratio $f_G/F_{\rm H\alpha}$
depends only on $\mathcal{R}$ which is only a function of $T_*$.
Notice that this ratio does not depend on the  (usually unknown)
distance to the nebula as well, as of course,  $R_*$. There are
two limitations to Equation~\ref{eq:f(G)}. We have assumed that the
nebula is substantial enough to consume all the ionizing photons.
If, instead, a fraction of the ionizing photons $\eta$ is used,
then $q\rightarrow \eta q$. Second, if the measured $F_{\rm H\alpha}$
is only from a portion of the nebula, say, $\zeta$, then the true
flux density of H$\alpha$ is $F_{\rm H\alpha}/\zeta$.  In either case,
$f_{\rm G}$ increases as $\eta^{-1}$ or $\zeta^{-1}$.

We normalize the flux density of H$\alpha$ to the limiting flux
density of the WHAM Point Source catalog, $f_{\rm H\alpha}\approx
10^{-11}\,{\rm erg\,cm^{-2}\,s^{-1}}$ and agree to use mJy for the
flux density of the G-band.  With these normalizations,
Equation~\ref{eq:f(G)} becomes 
 \begin{equation}
		f_{\rm G}({\rm mJy}) = 4.86\Big(\frac{f_{\rm
		H\alpha}}{\rm
		10^{-11}\,erg\,cm^{-2}\,s^{-1}}\Big)\mathcal{R}\ .
 \end{equation}
$\mathcal{R}$ varies from 1 ($T_*=3\times 10^4\,$K) to 0.03 ($T_*=10^5\,$K).
Thus, for a threshold object in WPS,   $f_{\rm G}$ correspondingly
ranges from 4.8\,mJy (G = 14.6\,mag) to  0.14\,mJy (G = 18.4\,mag).

\begin{figure}[htbp] 
 \centering
  \includegraphics[width=3in]{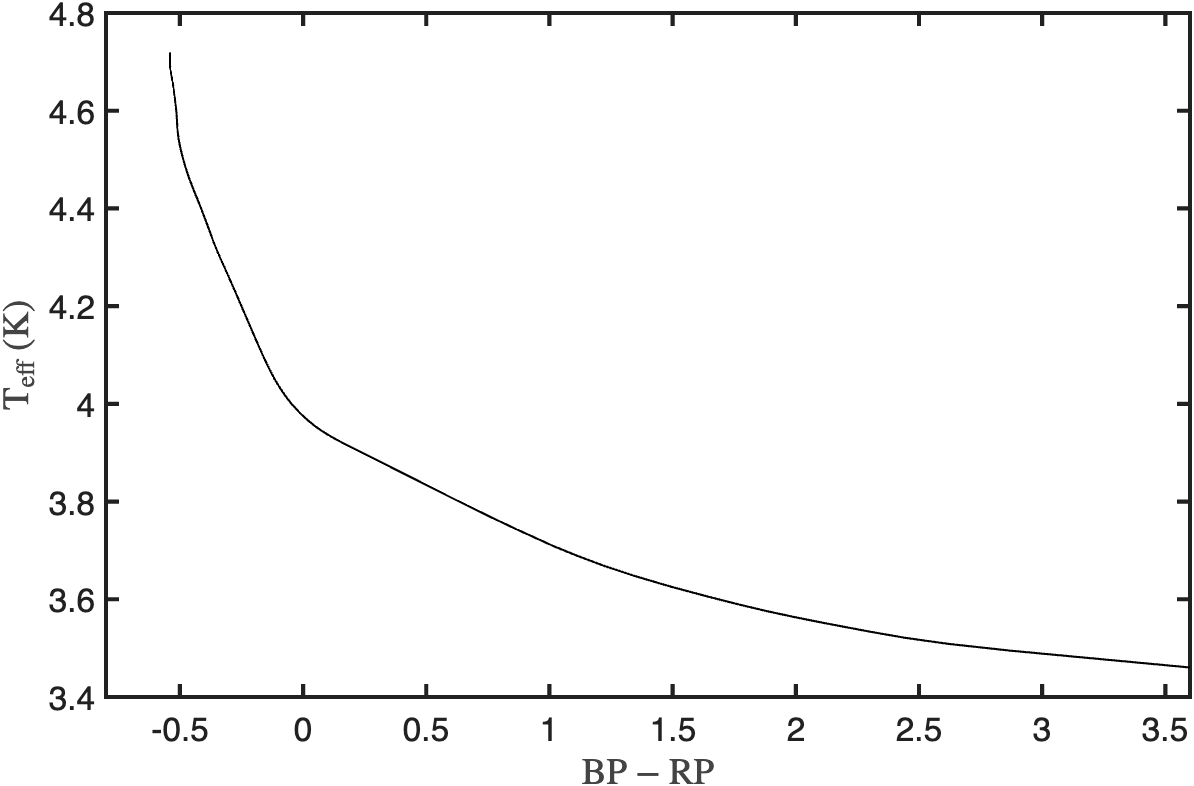}
   \caption{\small ${\rm BP-RP}$ versus effective temperature of
   stars. The relation was generated using MIST models for stars
   at the ZAMSs.  }
 \label{fig:Gaia_BPmRP_Teff} 
\end{figure}

Although $T_*$ is convenient from a theoretical perspective, it is more
convenient to use a color as a proxy for $T_*$. To this end, we use
${\rm BP-RP}$ which is conveniently available. Our adopted relation
of T$_*$ versus ${\rm BP-RP}$ is shown in
Figure~\ref{fig:Gaia_BPmRP_Teff}.

\section{PG\,0931+691}
	\label{sec:PG}

The star was observed by GALEX and SDSS and therefore enjoys good
photometric coverage (Table~\ref{tab:Photometry}). Hot sub dwarfs (sdO, sdB) have spectral similarity
to main sequence O/B stars but are sub-luminous (and thus smaller
radius). PG\,0931+691 is classified as He-sdO and
${\rm E(B-V) = 0.0971}$ \citep{gon+17}.
These are generally accepted to be core helium burning
stars (see the review article by \citealt{H16}).

\begin{deluxetable}{lrrr}[hbtp]   %PG0931_PhotTable.m
\label{tab:Photometry}
 \tablecaption{Photometric Data}
 \tablewidth{0pt}
 \tablehead{
 \colhead{band} &
 \colhead{mag}&
 \colhead{de-red} &
 \colhead{$f_\nu$}\cr
  \colhead{} & 
    \colhead{(mag)} &
 \colhead{(mag)} &
 \colhead{(mJy)}
 }
 {\small
 \startdata
 FUV & 15.176 & 14.701 & 4.78 \\
 NUV & 15.573 & 14.870 & 4.09 \\
   $u$ & 15.928 & 15.502 & 2.29 \\
   $g$ & 16.426 & 16.105 & 1.31 \\
   $r$ & 16.919 & 16.694 & 0.76 \\
   $i$ & 17.238 & 17.072 & 0.54 \\
   $z$ & 17.632 & 17.507 & 0.36 \\
\enddata
}
 \tablecomments{\small The first column is the band.  FUV and NUV
 are from GALEX while $ugriz$ is synthetic magnitude constructed
 from Gaia XP spectra \citep{vbj+24}.  The second column is the
 observed magnitude and the third column (``de-red") is magnitude
 corrected for reddening, E(B$-$V)= 0.0971 \citep{gon+17} and the
 final column is de-reddened spectral flux density.  There is
 excellent agreement between the synthetic Gaia photometry and
 $griz$ as measured by SDSS. See \cite{vbj+24} for a discussion
 related to $u$-band photometry.  }
\end{deluxetable}

\begin{figure}[htbp]   %PG0931_PhotTable.m,HotStarSpectralModel.m
\centering
 \includegraphics[height=1.7in]{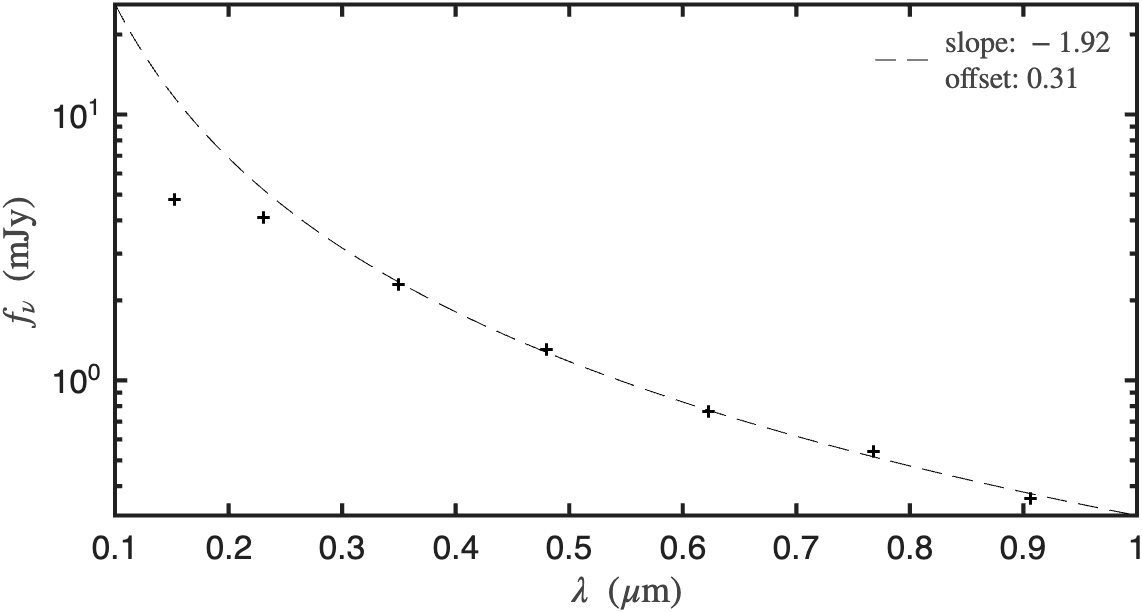}\qquad\qquad
 \includegraphics[height=1.7in]{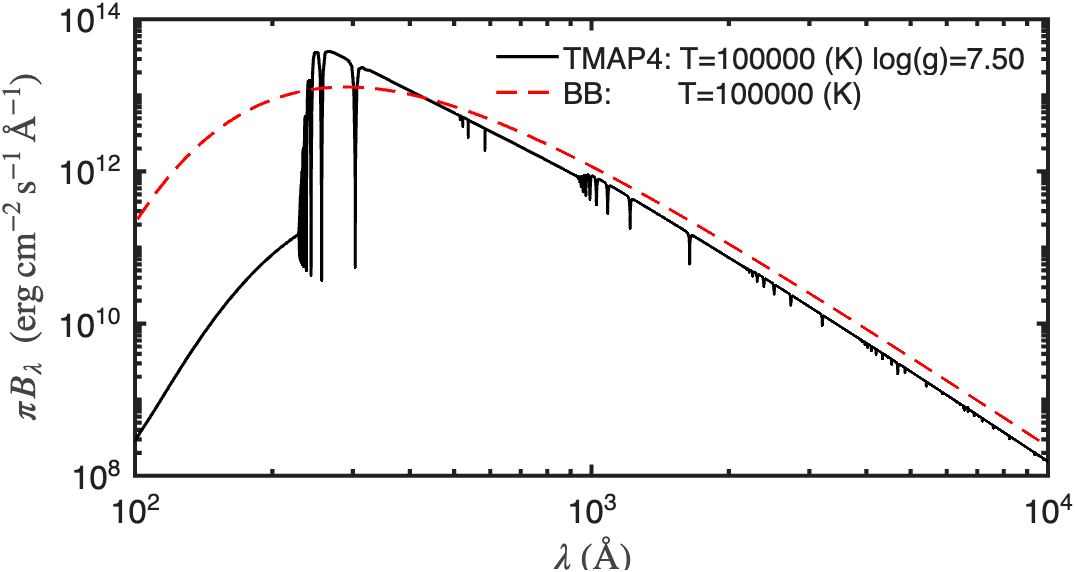}
  \caption{\small [Left]: The de-reddened photometry (``+"; in mJy)
  along with a  Rayleigh-Jeans model fit (dashed line):
  $f_\nu=a\lambda^{-2}$ where $f_\nu$ is in mJy, $\lambda$ in $\mu$m,
  $a=2k_BT\pi\theta^2$ and $\theta=(R_*/d)^2$.  [Right]:  The surface
  spectral flux, $\pi B_\lambda$, of a hot helium white dwarf (TMAP4;
  black line) and a star with blackbody emission (red dashed line).
  Here, $B_\lambda$ is the surface spectral intensity.  The physical
  parameters are shown in the legend. In the Rayleigh-Jeans tail
  of the spectrum the blackbody flux is 1.4 higher than the model
  flux. }
 \label{fig:Photometry_RJfit}
\end{figure}

We fit the reddened optical spectral flux density, $f_\nu$,
to a Rayleigh-Jeans model (Figure~\ref{fig:Photometry_RJfit}) and
find good agreement with this model, namely, $f_\nu\propto
\lambda^{-2}$. Thus, PG\,0931+691  is a very hot star.  The
Rayleigh-Jeans fit only yields $a\propto T(R_*/d)^2$ where $T$ is
the blackbody temperature and $R_*$ is the radius of PG\,0931+691.
The GALEX data are critical to determining the temperature.

In the Gaia XP catalog of white dwarfs, \cite{vbj+24} provide the
following model fit parameters:  $T_{\rm eff}=1.18\times 10^5\,$K,
$\log(g)=7.77$, inferred mass of $0.68\,M_\odot$ and helium atmosphere.
However, the Gaia  BP$-$RP of $-0.5$\,mag and $M_G$ of 8 mag places
the star in the region of hot white dwarfs rather than in the hot
sub-dwarf cloud. We learned from Dr.\ Nicole Reindl that this star
is an unusual DO white dwarf with helium atmosphere exhibiting
as-yet not understood ``Ultra-Highly Excited'' (UHE) metal line
emission.  

We obtained model theoretical spectra from the ``Theoretical
Spectra"\footnote{\url{https://svo2.cab.inta-csic.es/theory/newov2/}}
holdings of the Spanish Virtual Observatory (SVO), specifically the
TMAP4 collection \citep{rd03,wd+03}.  The radius of the white dwarf,
$R_*$, is a free parameter.  We also consider a simple blackbody
model (BB) with the same temperature.  In Figure~\ref{fig:Photometry_RJfit}
we present a TMAP4 stellar spectrum for a white dwarf with $T=10^5\,$K,
$\log(g)=7.5$ and helium atmosphere along with that from a black-body
of the same temperature.

For each model (TMAP, BB) we compute $q_s$, the rate of photons capable
of ionizing hydrogen per cm$^2$ of the surface. Next, we compute
$f_\nu(r)$, the flux in the $r$ band assuming that the white dwarf
is located at $d=580\,$pc.  We adjust $R_*$ so that the $r$ band
flux is approximately proportionate to the reddened measured flux
(Table~\ref{tab:Photometry}).  Finally, equation $Q=4\pi R_*^2q_s$
yields the total ionizing rate for each model.  

\begin{deluxetable}{lrrrrr}[hbtp]    %
\label{tab:Q_fnu}
 \tablecaption{sDO white dwarfs}
 \tablewidth{0pt}
 \tablehead{
 \colhead{model} &
 \colhead{$T$} & 
 \colhead{$R_*$} & 
 \colhead{$q_{26}$}&
  \colhead{$Q_{45}$}&
 \colhead{$f_\nu(r)$}
 }
 {\small
 \startdata
TMAP & $10\times 10^4$ & 0.017 & 1.04 & 1.83 & 0.560 \\
BB        & "                           & "        & 1.07 & 1.88 & 0.867 \\
\hline
TMAP & $8\times 10^4$ & 0.020 & 0.438 & 1.06 & 0.634 \\
BB      & "                        & "         & 0.465  & 1.13 & 0.931
\enddata
}
 \tablecomments{\small  The temperature and radius are given in
 columns 2 and 3.  $q_s=10^{26}q_{26}\,{\rm cm^{-2}\,s^{-1}}$ and
 $Q=10^{45}Q_{45}\,{\rm s^{-1}}$ are given in columns 4 and 5. The
 last column is $f_\nu(r)$, the model $r$ band flux in mJy.
 $\log(g)=7.5$ for both TMAP models. A distance of 580\,pc is
 assumed. }
\end{deluxetable}

\section{Imaging the Vulcan Nebula}
 \label{sec:Vulcan}

The sky region in the vicinity of WPS\,46 was imaged during the
period November 2024 and June 2025 by the following group of amateur
astronomers: Marty Anderson, Keith Mombourquette, Mike Tettenborn
and Adam Hofmann.

M.\ Anderson imaged the field with a Samyang 135 telescope (stopped at $f$/2.5)
and an ASI2600MC OSC camera with UV/IR filter and filterless for
a total of 40 hours and one hour  filterless but with $f$/10. 
Anderson
made H$\alpha$ (3\,nm bandwidth) observations with a Recat71
Petzval APO refractor (350\,mm $f$/4.9) coupled to an ASI2600MM
camera for a period of 37 hours. K.\ Mombourquette imaged in the [\ion{O}{3}]
3-nm wide filter for a total of 60.5 hours with an Askar FRA300 and
107PHQ telescope coupled to ASI2600MM cameras.  With an Askar
107PHQ telescope coupled to an ASI2600MM camera, M.  Tettenborn
imaged M8/M82 for 8 hours in LGRB-V filters (2 hours of each) and
5.5 hours (Antlia 3-nm H$\alpha$ filter). A.\ Hofmann used an Askar
ACL200 (200 mm; $f$/4) coupled to a ZWO ASI 2600MM mono camera to
image the region in H$\alpha$ (4.5-nm bandwidth) for a total of 33
hours.

Data processing was carried out by M. Anderson using Pixinsight and
Affinity Photo, with K.\ Mombouquette providing support. In 2025,
the North York Astronomical Association (NYAA) introduced a new
category of recognition (“The DSO Collaboration category was added
to encourage teams of up to four people to combine data collected
over the last 10 years to create super-long integrations.”) The
image presented in Figure~\ref{fig:Vulcan} won the ``DSO Collaboration''
award in 2025.

\end{document}